\documentclass[pdflatex,sn-mathphys-num]{sn-jnl}

\usepackage{subfigure}
\usepackage{graphicx}%
\usepackage{multirow}%
\usepackage{amsmath,amssymb,amsfonts}%
\usepackage{amsthm}%
\usepackage{mathrsfs}%
\usepackage[title]{appendix}%
\usepackage{xcolor}%
\usepackage{textcomp}%
\usepackage{manyfoot}%
\usepackage{booktabs}%
\usepackage{algorithm}%
\usepackage{algorithmicx}%
\usepackage{algpseudocode}%
\usepackage{listings}%

\theoremstyle{thmstyleone}%
\theoremstyle{thmstyletwo}%

\theoremstyle{thmstylethree}%

\begin{document}

\title[Bursty Switching Dynamics Promotes the Collapse of Network Topologies]{Bursty Switching Dynamics Promotes the Collapse of Network Topologies}


\author[1]{\fnm{Ziyan} \sur{Zeng}}

\author*[1]{\fnm{Minyu} \sur{Feng}}\email{myfeng@swu.edu.cn}

\author*[2,3,4,5,6]{\fnm{Matja\v{z}} \sur{Perc}}\email{Matjaz.Perc@gmail.com}

\author[7,8]{\fnm{J\"{u}rgen} \sur{Kurths}}

\affil[1]{College of Artificial Intelligence, Southwest University, 400715, Chongqing, China}

\affil[2]{Faculty of Natural Sciences and Mathematics, University of Maribor,
Koro{\v s}ka cesta 160, 2000 Maribor, Slovenia}

\affil[3]{Community Healthcare Center Dr. Adolf Drolc Maribor, Ulica talcev 9,
2000 Maribor, Slovenia }

\affil[4]{Department of Physics, Kyung Hee University, 26 Kyungheedae-ro,
Dongdaemun-gu, Seoul 02447, Republic of Korea}

\affil[5]{Complexity Science Hub, Metternichgasse 8, 1030 Vienna, Austria}

\affil[6]{University College, Korea University, 145 Anam-ro, Seongbuk-gu,
Seoul 02841, Republic of Korea}

\affil[7]{Department of Complexity Science, Potsdam Institute for Climate Impact Research, Potsdam, Germany }

\affil[8]{Institute of Physics, Humboldt University of Berlin, Berlin, Germany}


\abstract{Time-varying connections are crucial in understanding the structures and dynamics of complex networks. In this paper, we propose a continuous-time switching topology model for temporal networks that is driven by bursty behavior and study the effects on network structure and dynamic processes. Each edge can switch between an active and a dormant state, leading to intermittent activation patterns that are characterized by a renewal process. We analyze the stationarity of the network activation scale and emerging degree distributions by means of the Markov chain theory. We show that switching dynamics can promote the collapse of network topologies by reducing heterogeneities and forming isolated components in the underlying network. Our results indicate that switching topologies can significantly influence random walks in different networks and promote cooperation in donation games. Our research thus provides a simple quantitative framework to study network dynamics with temporal and intermittent interactions across social and technological networks.}




\maketitle
\section{Introduction}
Network theory serves as a comprehensive framework for describing the interconnections among various entities \cite{newman2002random, bollobas2013modern}. The discovery of "small-world" \cite{watts1998collective} and "scale-free" \cite{barabasi1999emergence} networks has catalyzed significant interest in the formation and statistical properties of numerous real-world complex networks \cite{fernex2021cluster, serafino2021true, sekara2016fundamental,bontorin2024emergence}, such as the Internet and collaboration networks. These seminal studies have elucidated phenomena such as the six degrees of separation and the power-law degree distribution observed in large-scale networks.

In the past two decades, research has primarily concentrated on network modeling to investigate how various mechanisms affect network topologies, often based on real-world system assumptions, such as fitness \cite{zhou2023nature, bianconi2001bose} or variable increments \cite{feng2020accumulative}. As the field of complex networks has evolved, increasing attention has been directed toward time-varying relationships among individuals \cite{holme2012temporal, williams2022shape, li2017fundamental}, including human contact \cite{barrat2013empirical}, communication \cite{estrada2013communicability,carmody2022effect}, and protein interactions \cite{taylor2009dynamic}. The concept of temporal networks offers a novel perspective on understanding the evolving structures of networks and their dynamic processes. In temporal networks, both vertex and edge sets are assigned timestamps to represent the continually changing interactions within a population. Utilizing this temporal network framework, researchers have achieved significant insights into dynamic processes, including epidemic propagation \cite{perra2012activity, valdano2018epidemic,van2021theory} and evolutionary dynamics \cite{li2020evolution, 10521680,10269140}. 

One method for investigating the effects of temporal links in complex networks involves examining a switching topology \cite{olfati2004consensus, ning2017collective}. Existing literature often models network structure changes as governed by a switching signal \cite{saboori2014h_, zhang2017distributed}, where switching events occur at predetermined times specified by the signal, leading to corresponding alterations in network connections \cite{wells2015control}. In this paper, we focus on a more generalized framework where the timing of topology switches is determined by the intermittent bursty interactions among entities \cite{sheng2023constructing, lee2018hierarchical}. 

Interactions in many time-varying complex systems typically exhibit bursty behavior, characterized by individuals contacting their peers—such as sending emails or making phone calls—based on immediate needs. Conventional models often assume that these interactions follow Poisson statistics \cite{miritello2011dynamical, zeng2023temporal}, which results in an exponential distribution of interevent times and implies that the number of events is a linear function of time. However, recent empirical studies have demonstrated that interevent times in complex systems often adhere to non-Poisson statistics, typically exhibiting heavy-tailed distributions \cite{barabasi2005origin, ross2015understanding,bagrow2013natural}. This suggests that events do not occur uniformly over time. Such non-Poisson intermittent activation patterns are observed in various real-world systems, including email communication \cite{malmgren2008poissonian} and neuronal activity \cite{kemuriyama2010power}. Mathematical modeling of bursty behavior is crucial for understanding the properties and dynamics of complex systems, including population models \cite{dos2022metapopulation, hiraoka2020modeling} and dynamics \cite{masuda2020small, han2023impact, yao2023inhibition}. 

In this article, we investigate the switching topology induced by bursty behaviors across a range of distributions, encompassing both Poisson and non-Poisson statistics. We model the switching topology of temporal networks by considering that each edge can alternate between activated and dormant states within the underlying static network. On the continuous time axis, both the activation and dormancy periods of an edge are modeled as random variables following two independent general distributions, resulting in a cyclic but not strictly periodic renewal process. This framework allows the switching signal for an edge to transition between activated and dormant states based on random timing, applicable to both Poisson and non-Poisson bursty behaviors. We conduct a theoretical analysis of the impact of this switching topology on network structure and explore its effects on collective behaviors by examining random walks and the prisoner's dilemma within networks exhibiting stochastic switching topologies.
\section{Results}\label{sec2}
\begin{figure}
    \centering
    \includegraphics[scale=0.06]{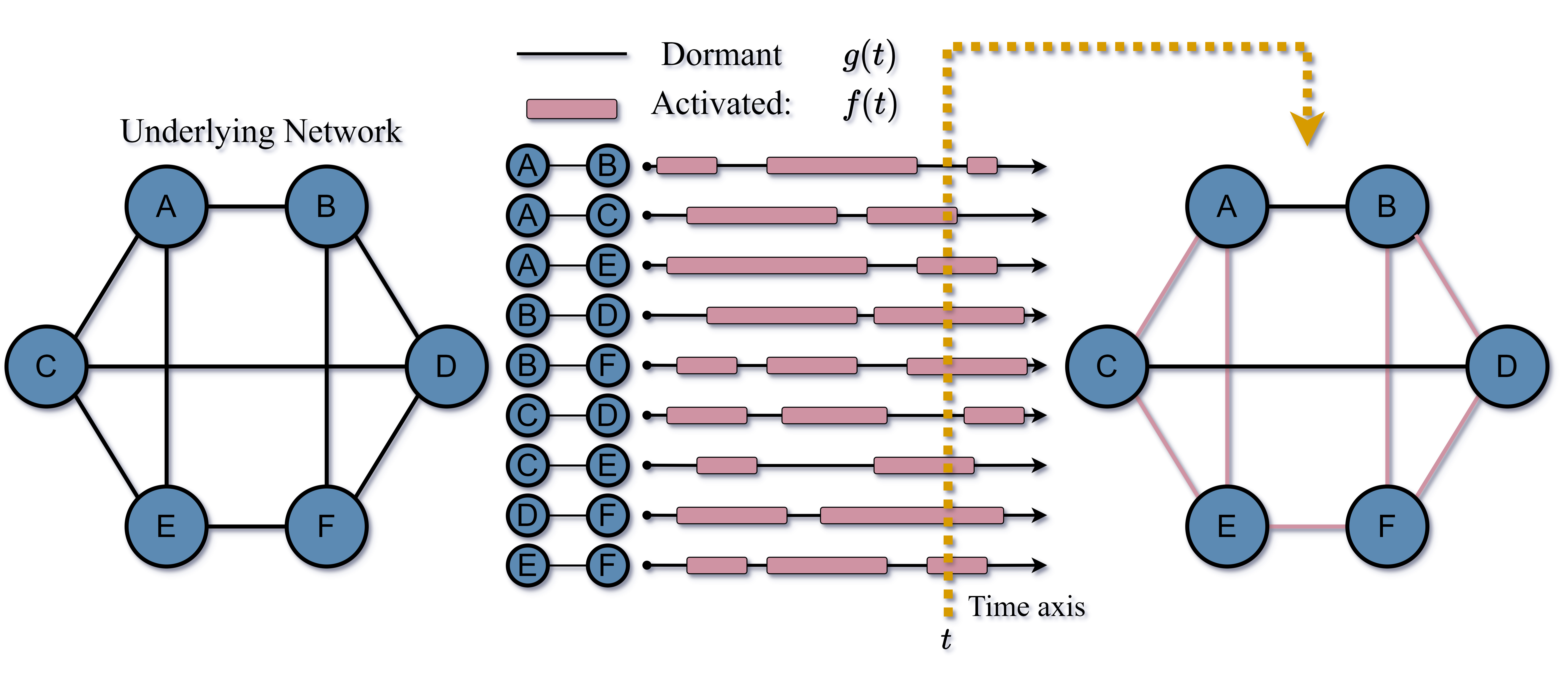}
    \vspace{-5mm}
    \caption{\textbf{Illustration of stochastic edge-switching topology in complex network. }The underlying network (left panel) is static with 6 vertices and 9 edges. Each edge switches between dormant and activated states independently in a continuous time axis (middle panel). The time spent in activated and dormant states is governed by general distributions $f(t)$ and $g(t)$. At the time $t$, the edge set of activated subgraph $\mathcal{G}_A(t)$ does not contain the dormant edges $(A, B)$ and $(C, D)$. }
    \label{fig: network example}
\end{figure}
We present an edge-switching dynamic model for complex networks, where each edge can exist in one of two states: activated or dormant. The duration for which an edge remains in either state is governed by random variables that follow two independent probability distributions, which may be either Poisson or non-Poisson in nature. We employ continuous-time Markov chain theory to analyze the effects of these dynamic state transitions on the network's topological properties. Building on our analysis of the network's structural characteristics, we further investigate the implications of the proposed switching topology on random walks and evolutionary dynamics within the network.

\subsection{Switching Topology Model with Bursty Dynamics}
We consider an undirected and unweighted static network $\mathcal{G}$ as the underlying network, with the vertex set $\mathcal{V}$ ($\vert \mathcal{V} \vert= N$), the edge set $\mathcal{E}$ ($\vert \mathcal{E}\vert= M$), and the degree distribution ($p(k)$, the probability to find vertex with degree $k$ in the whole network). The average degree of the focal underlying network $\mathcal{G}$ is $\left\langle k\right\rangle=\sum_0^\infty kp(k)$. We assume that each edge is in one of the activated and dormant states. The edge activation is a non-instantaneous event. Once an edge turns activated at a moment, it turns dormant after a random time that follows the probability distribution $f(t)$. Similarly, once an edge turns dormant, it is activated again after another random time that follows the distribution $g(t)$. We define the activated subgraph at time $t$ as $\mathcal{G}_A(t)=(\mathcal{V}, \mathcal{E}_A(t))$, where $\mathcal{E}_A(t)$ is the activated edge set at time $t$. In the following analysis, we mainly focus on the property of the activated subgraph because the edge set of the dormant subgraph is $\mathcal{E}_D(t)=\mathcal{E}-\mathcal{E}_A(t)$ and hence has similar properties. Additionally, a vertex is considered "trapped" if all edges incident to it are dormant at a given time.  We present an example of this model in Fig.~\ref{fig: network example}. 
\begin{figure}
\centering
\subfigure[$m$ vs. Time]{\includegraphics[scale=0.23]{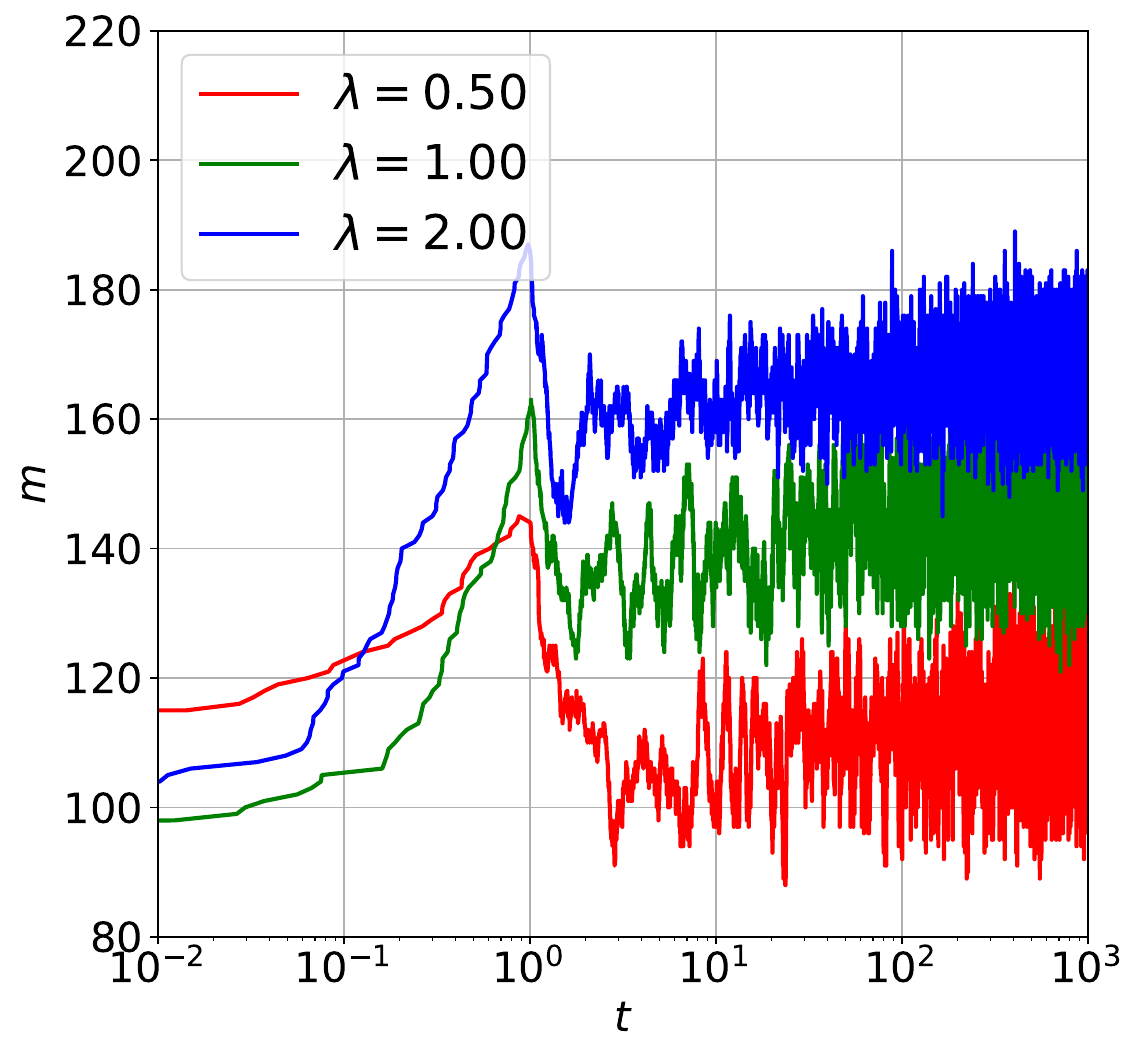}}\label{fig: edge number time}
\hspace{-3mm}
\subfigure[Stationary Distribution of $m$]{\includegraphics[scale=0.23]{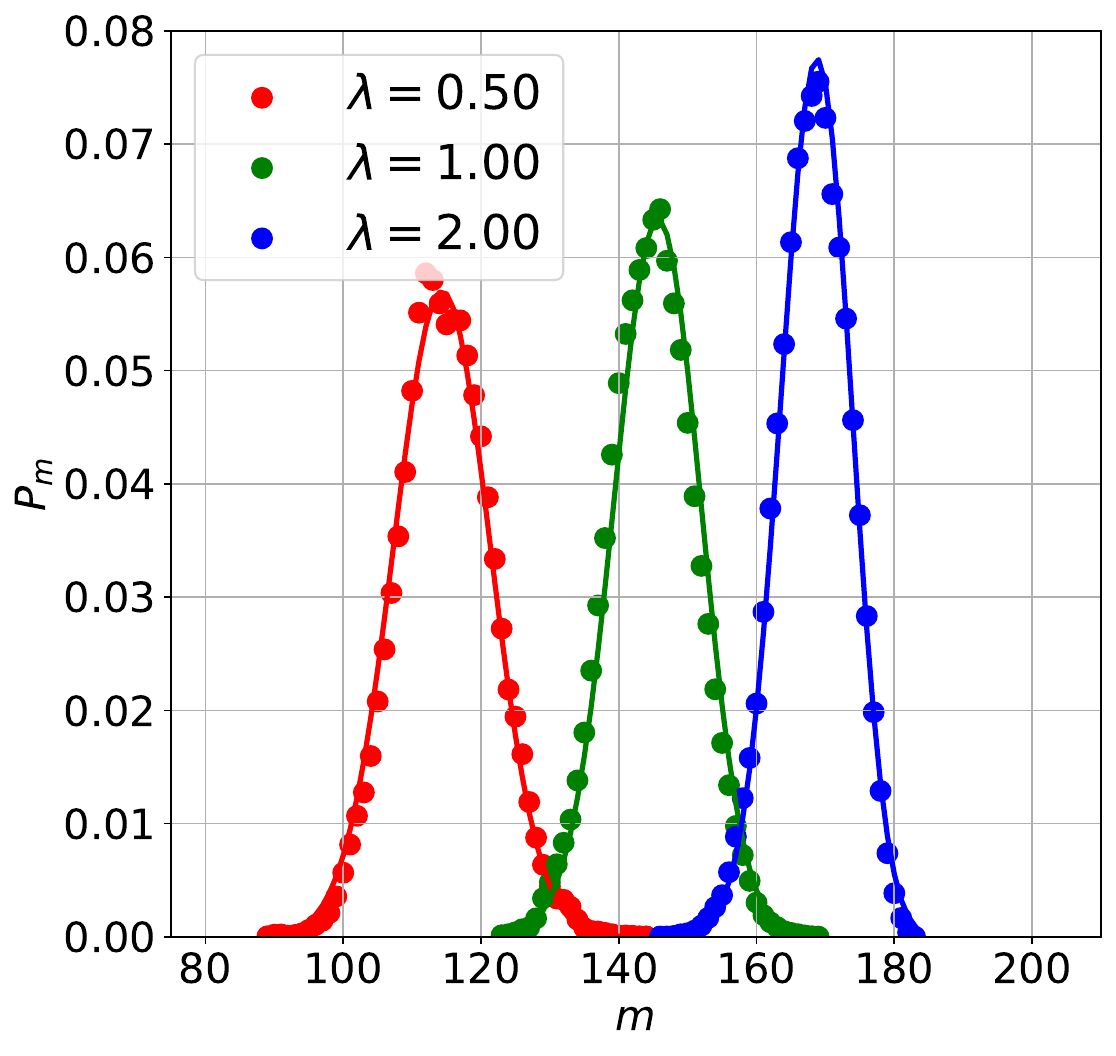}}\label{fig: edge number}
\hspace{-2mm}
\subfigure[Degree Distribution]{\includegraphics[scale=0.23]{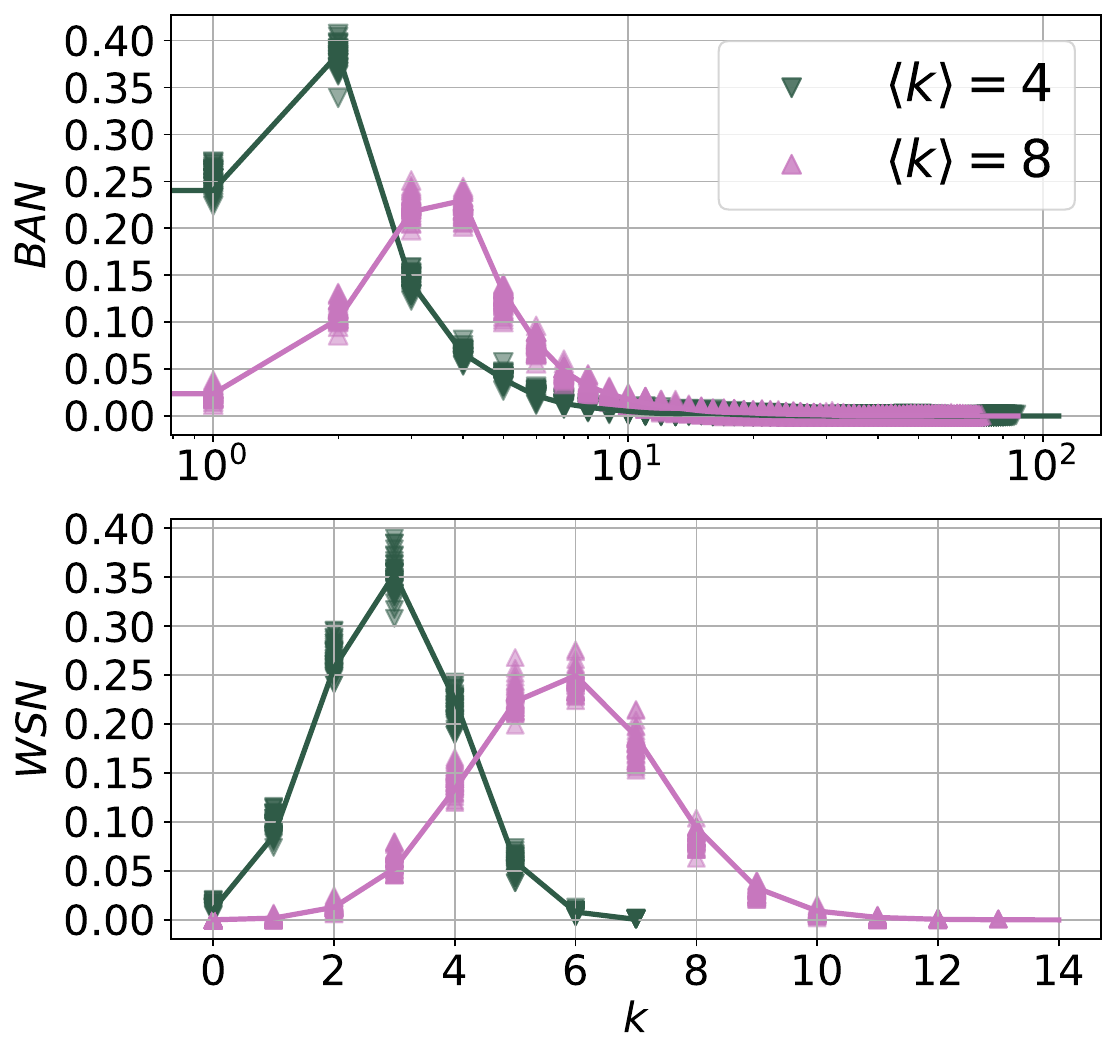}}\label{fig: degree distribution}
\hspace{-2mm}
\vspace{-3mm}

\subfigure[Density]{\includegraphics[scale=0.23]{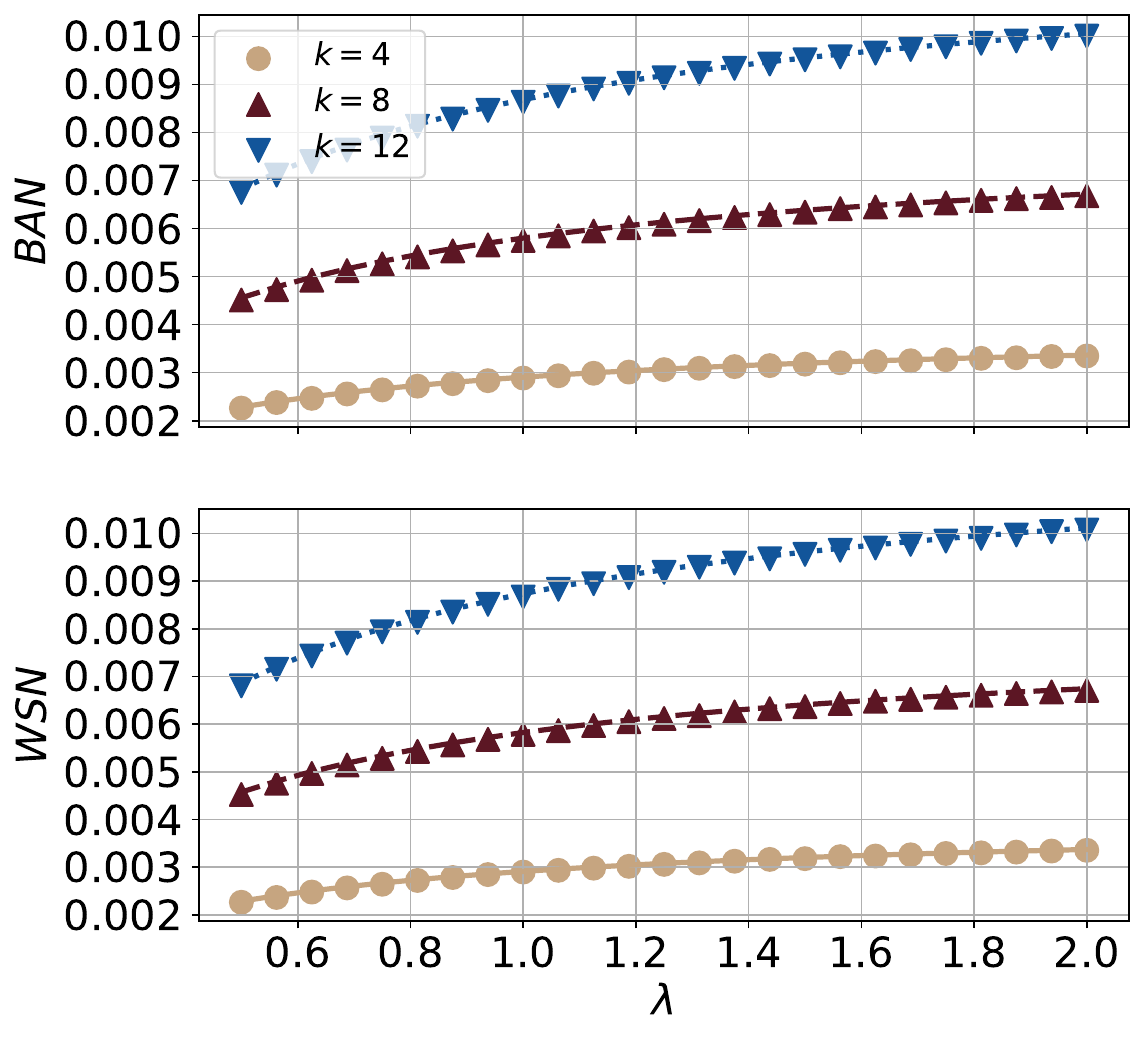}}\label{fig: density}
\hspace{-2mm}
\subfigure[Spectral Radius]{\includegraphics[scale=0.22]{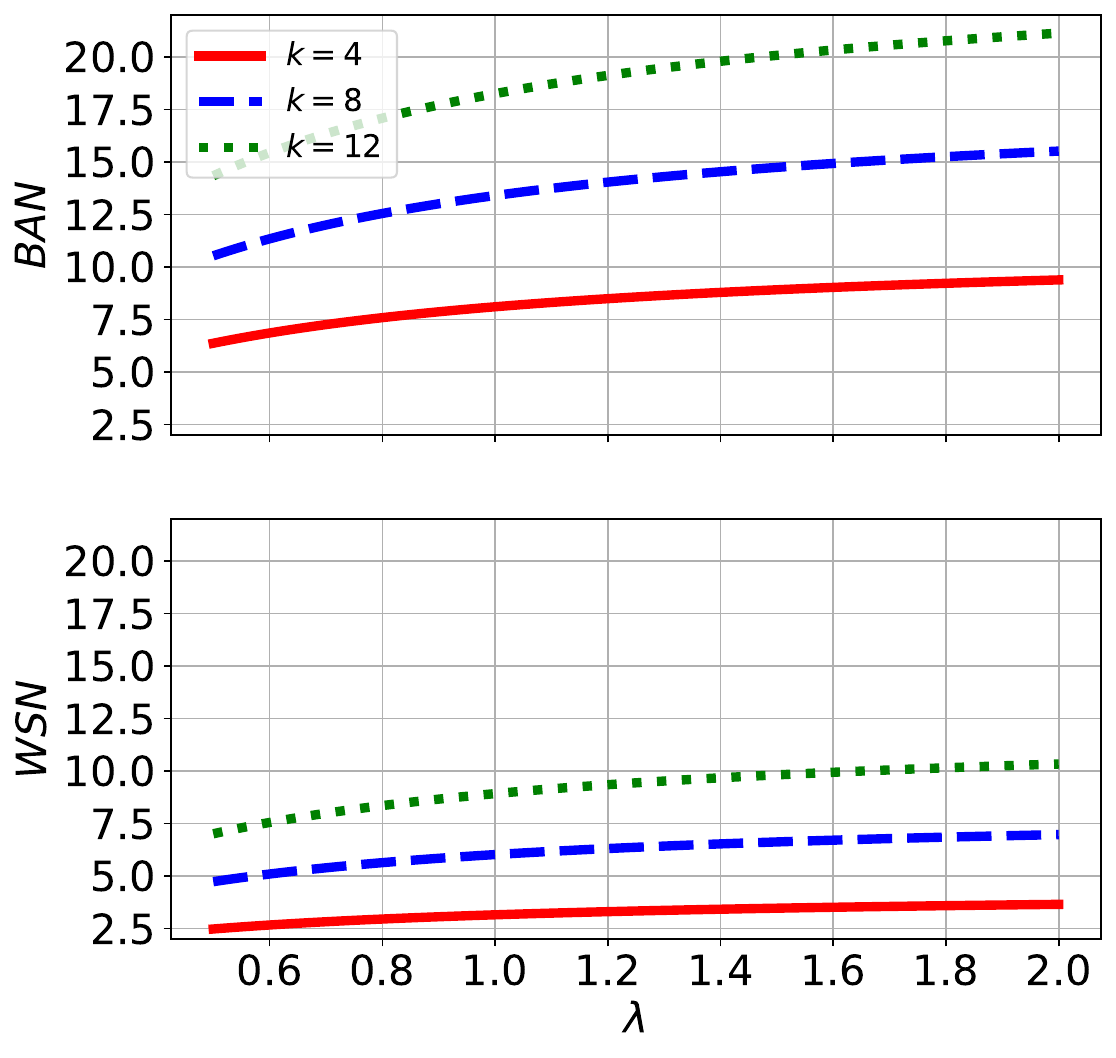}}\label{fig: spectral radius}
\hspace{-2mm}
\subfigure[Maximum Component]{\includegraphics[scale=0.23]{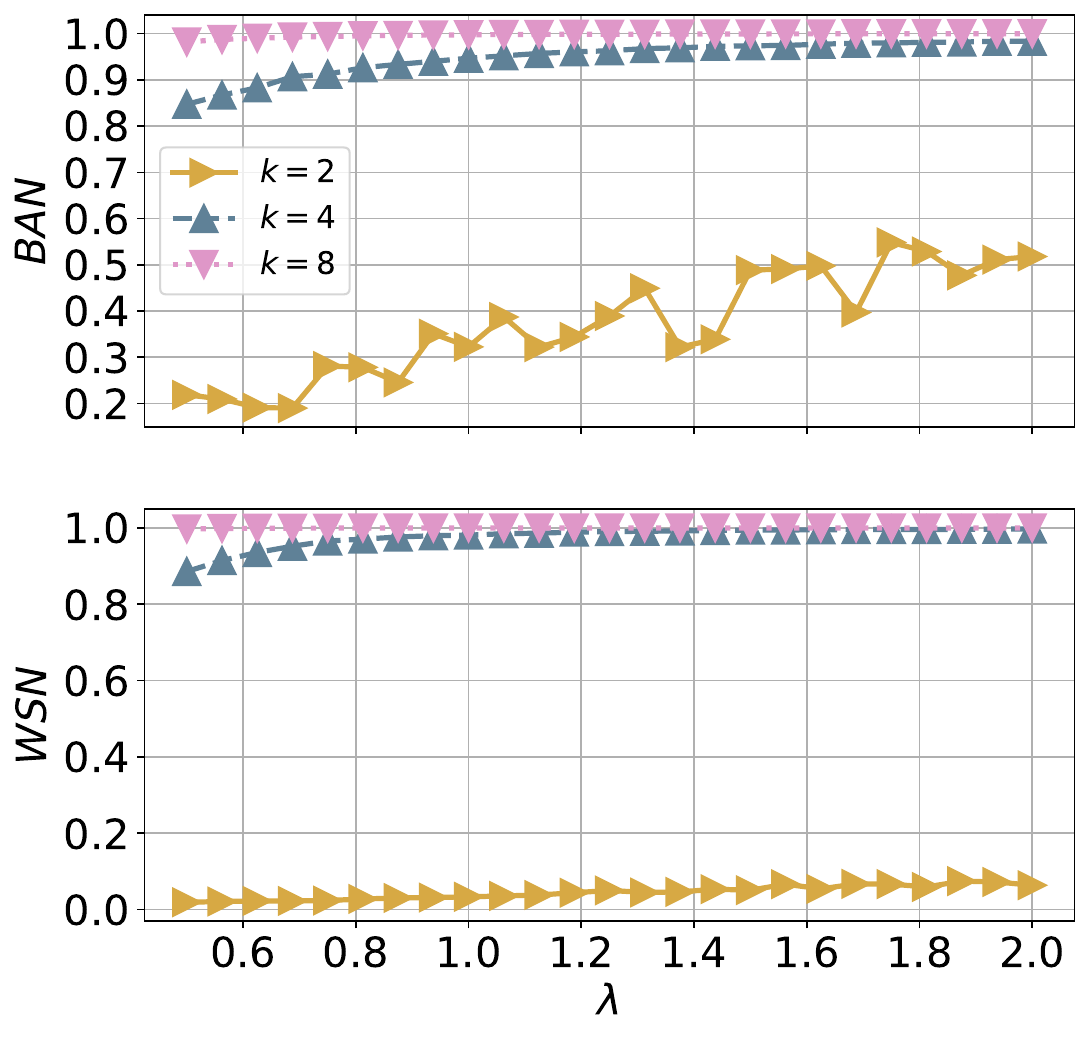}}
\vspace{-3mm}
\caption{\textbf{Network topology properties. } (a) Evolution of activated edge numbers with time. The underlying network is RRG with $N=100$ and $\left\langle k\right\rangle=4$. We set $\lambda\in\{0.50, 1.00, 2.00\}$ and $\alpha=2.60$ with randomly assigned initial condition. (b) Stationary distribution of activated edge number. The network and switching topology settings are the same as (a). Each data point is averaged in the stable state with $t\in[0.5\times10^3,10^3]$. Theoretical results are shown in curves with corresponding colors. (c) Degree distribution of activated subgraph $\mathcal{G}_A(t)$. The underlying networks are BAN (upper panel) and WSN (lower panel) with $N=10^3$ and $k\in\{4,8\}$. The switching parameters are $\lambda=1$ and $\alpha=2.6$. We plot the numerical degree distributions in data points in $t\in[50,100]$ with the interval 2. Curves are the theoretical degree distribution. (d) Density of $\mathcal{G}_A(t)$. The underlying networks are BAN (upper panel) and WSN (lower panel) with $N=2\times10^3$ and $k\in\{4,8,12\}$. The switching parameters are $\lambda\in[0.5,2.0]$ and $\alpha=2.6$. Each data point is the average of over 100 independent realizations in $t\in[100,300]$. Theoretical results are shown in curves. (e) The spectral radius of the average adjacency matrix of $\mathcal{G}_A(t)$. The network and switching topology settings are the same as (d). (f) The relative size of the largest component. The underlying networks are BAN (upper panel) and WSN (lower panel) with $N=10^3$ and $k\in\{2,4,8\}$. The switching parameters are the same as (d) and (e). Each data point is the average of over 10 independent realizations in $t\in[50,150]$. For (c)-(f), the reconnection probability of WSNs is unified as $0.25$. }\label{fig: network topology}
\end{figure}

We are interested in the network property with the proposed switching topology mechanism. We mainly focus on the network topology properties averaged over time (expected quantities). Concretely, we pay attention to the expected topological property of the activated subgraph $\mathcal{G}_A(t)$. We first consider one single edge and denote the quantity $x_{ij}(t)\in\{0, 1\}$ as the state of the edge $(i,j)$ at time $t$, where $0$ and $1$ indicate that this focal edge is dormant or activated respectively. If $i$ and $j$ are not connected in the $\mathcal{G}$, we have $x_{ij}(t)=0$ for an arbitrary time. Accordingly, the activated edge number can be expressed by $\sum_{i,j\in\mathcal{G}}x_{ij}(t)$. The evolution of this quantity can be regarded as a queueing system with a limited source. For the connected pair $i$ and $j$, $x_{ij}(t)$ is a renewal process that takes values in $\{0, 1\}$ by turns. If the process $x_{ij}(t)$ has completely gone through two adjacent activated and dormant states during time $T_C$, then we say this process undergoes a \textit{cycle}. The expected time for a cycle is
\begin{equation}
E[T_C]=\int_0^{\infty} t[f(t)+g(t)]dt. 
\end{equation}
We define the activation constant as $q_0=\int_0^{\infty} tf(t)dt/E[T_C]$, i.e., the probability of finding an edge activated in a cycle. Since we assume the activated and dormant states of all edges are independent, the probability of finding $m$ activated edges in the underlying network $\mathcal{G}$ follows the binomial form
\begin{equation}\label{eq: binomial}
P[\sum_{i,j\in\mathcal{G}}x_{ij}(t)=m]=\frac{M!}{m!(M-m)!}q_0^m (1-q_0)^{(M-m)}. 
\end{equation}

Therefore, the density of $\mathcal{G}_A(t)$ is $E[\rho]=2q_0M/N(N-1)$. We note that a vertex $i$ can have no activated edge and becomes isolated in $\mathcal{G}_A(t)$, with the trapped probability $P(\sum_{j\in\mathcal{G}}x_{ij}(t)=0)=(1-q_0)^{k_i}$, where $k_i$ is the vertex $i$'s degree in the underlying network. We are also concerned about the expected degree distribution of $\mathcal{G}_A(t)$, i.e., the probability of finding a vertex with degree $k$ in the activated subgraph. With the proposed phase transition of edge states, the degree distribution of the activated subgraph ($\mathcal{G}_A(t)$) of $\mathcal{G}$ is determined by $p(k)$ and $q_0$. Suppose that we find an individual with the degree $j$ in the network $\mathcal{G}$ as the focal vertex. With the state transitions of the edges connected to its neighbors, its degree in $\mathcal{G}_A(t)$ varies as an integer in the range $[0, j]$. Obviously, at a sufficient time $t$, this focal vertex has the degree $i$ with the probability $\binom{j}{i} q_0^i (1-q_0)^{(j-i)}$. We have mentioned that the degree distribution of the network $\mathcal{G}$ is $p(j)$, i.e., the probability of finding a vertex with degree $j$. Consequently, the expected degree distribution of $\mathcal{G}A(t)$ can be expressed as (see Methods \ref{Methods: 1} for details)
\begin{equation}\label{eq: degree distribution}
p_A(i)=\sum_{j \geq i} \frac{j!}{i!(j-i)!} q_0^i (1-q_0)^{j-i} p(j). 
\end{equation}
\begin{figure*}                                 
\centering
\subfigure[Yeast]{\includegraphics[scale=0.24]{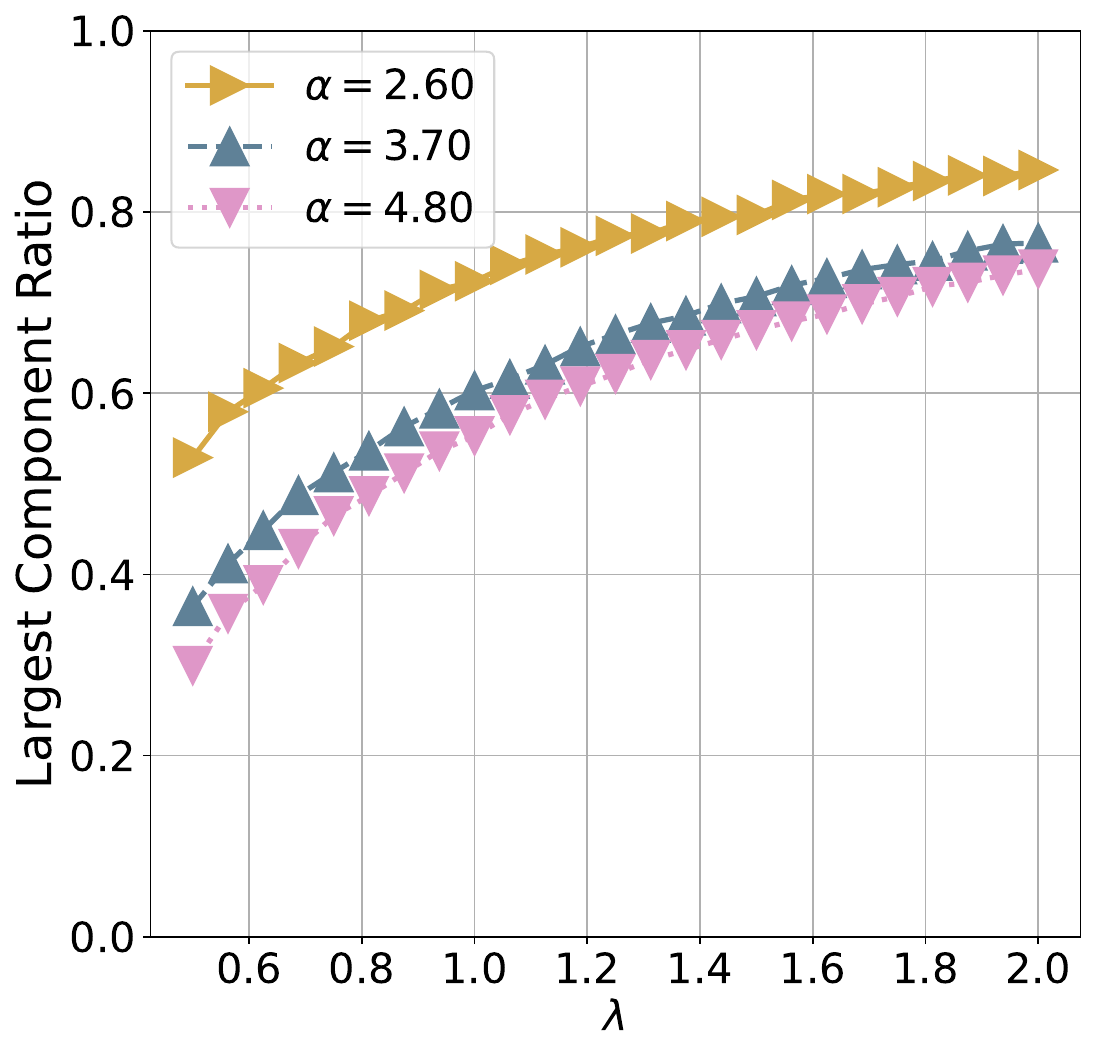}}\label{fig: component bio-yeast}
\hspace{-2mm}
\subfigure[WormNet-CE-HT]{\includegraphics[scale=0.24]{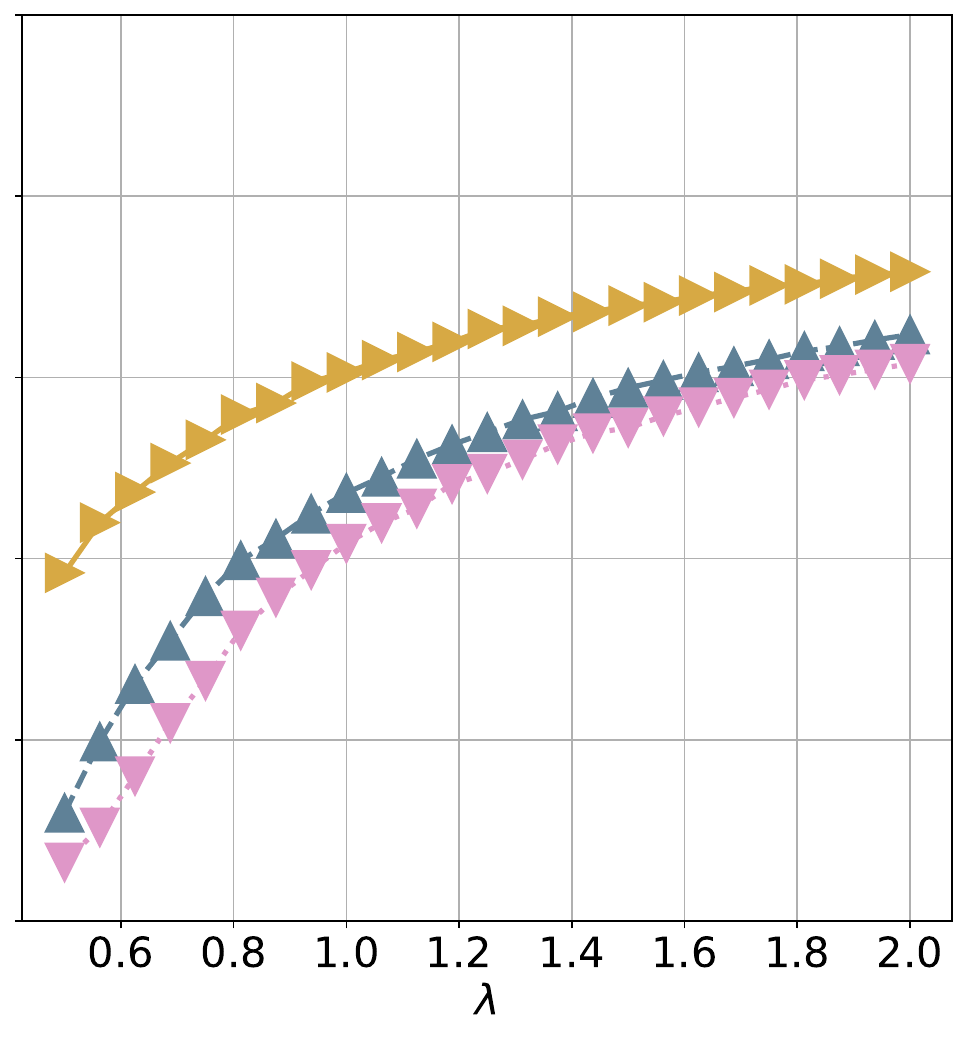}}\label{fig: component bio-CE-HT}
\hspace{-2mm}
\subfigure[WormNet-CE-LC]{\includegraphics[scale=0.24]{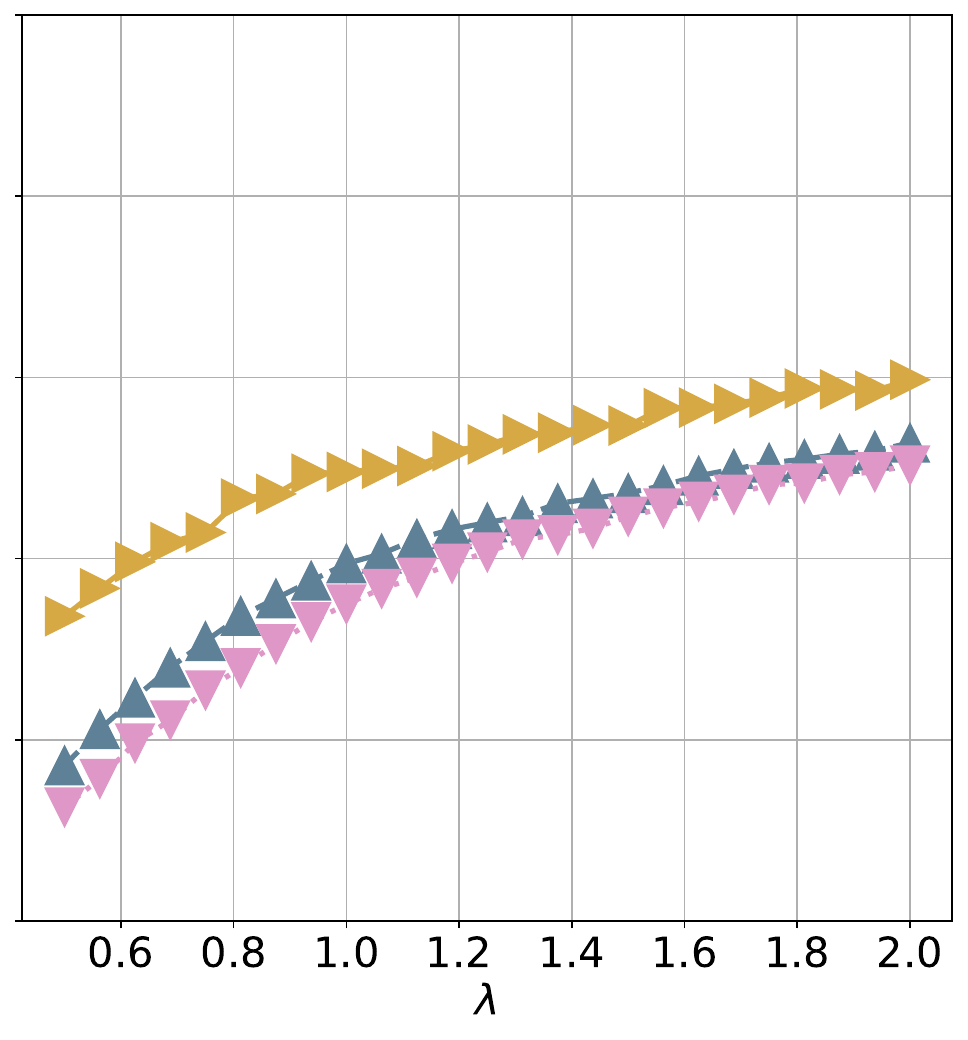}}\label{fig: component bio-CE-LC}
\hspace{-2mm}

\subfigure[Moreno Crime]{\includegraphics[scale=0.243]{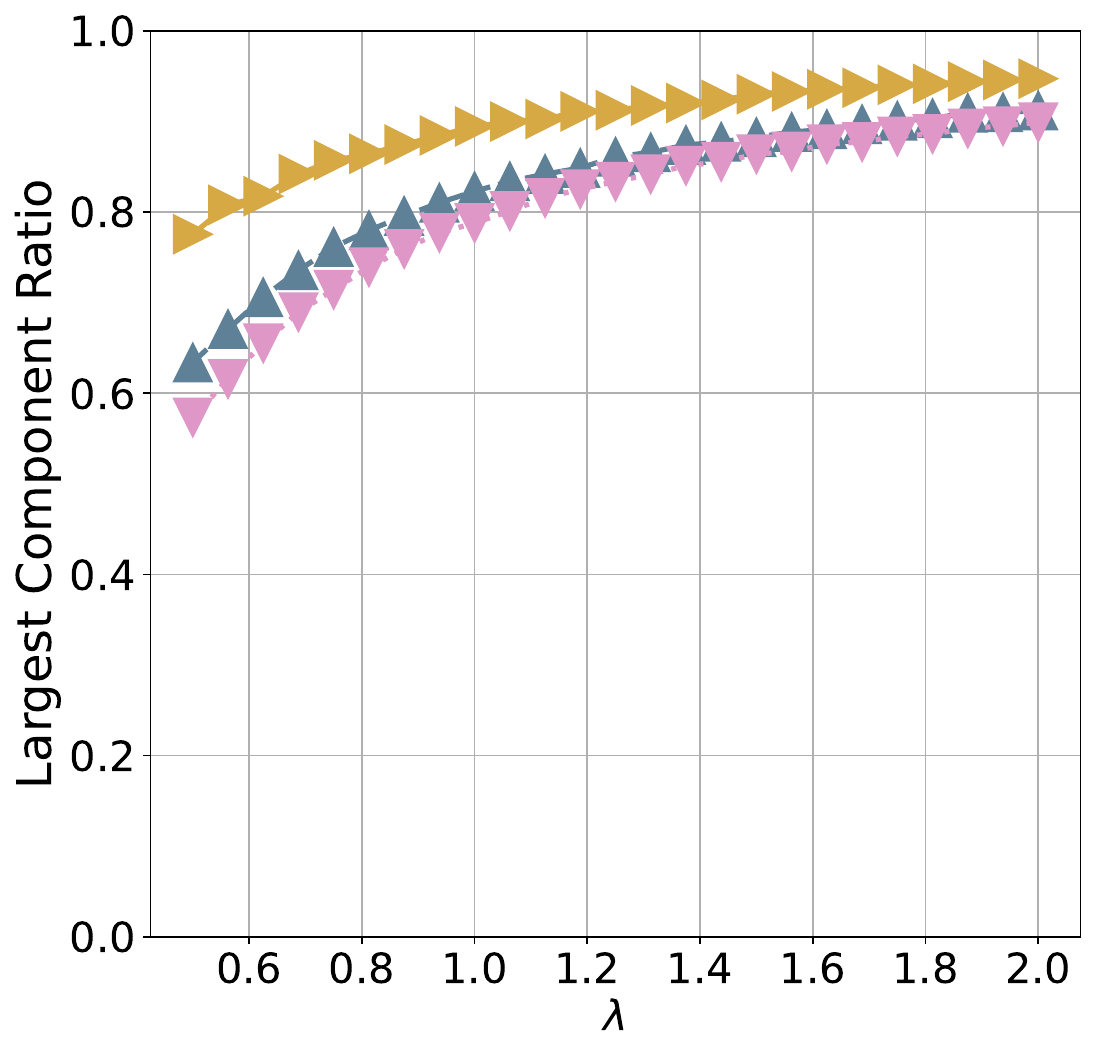}}\label{fig: component ia-crime-moreno}
\hspace{-2mm}
\subfigure[Brain]{\includegraphics[scale=0.24]{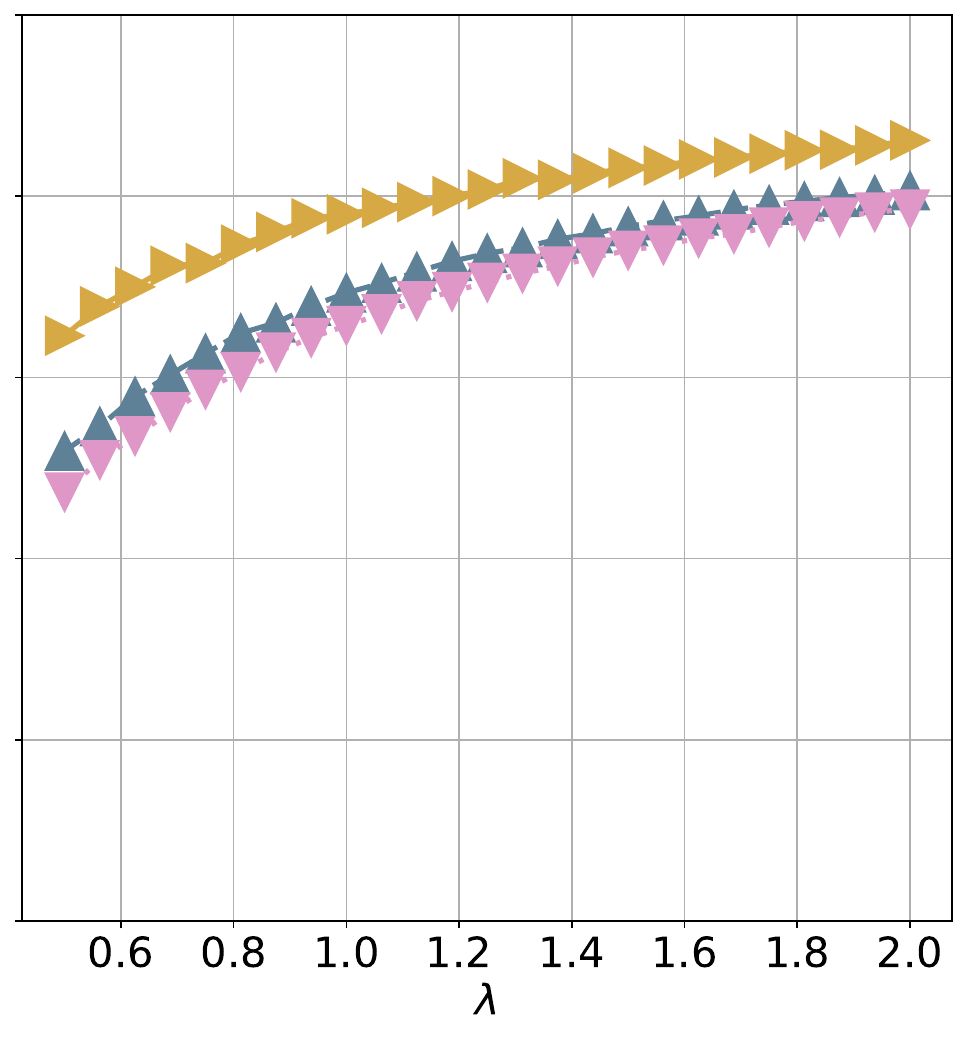}}\label{fig: component bn-mouse-kasthuri_graph_v4}
\hspace{-2mm}
\subfigure[Retweet]{\includegraphics[scale=0.24]{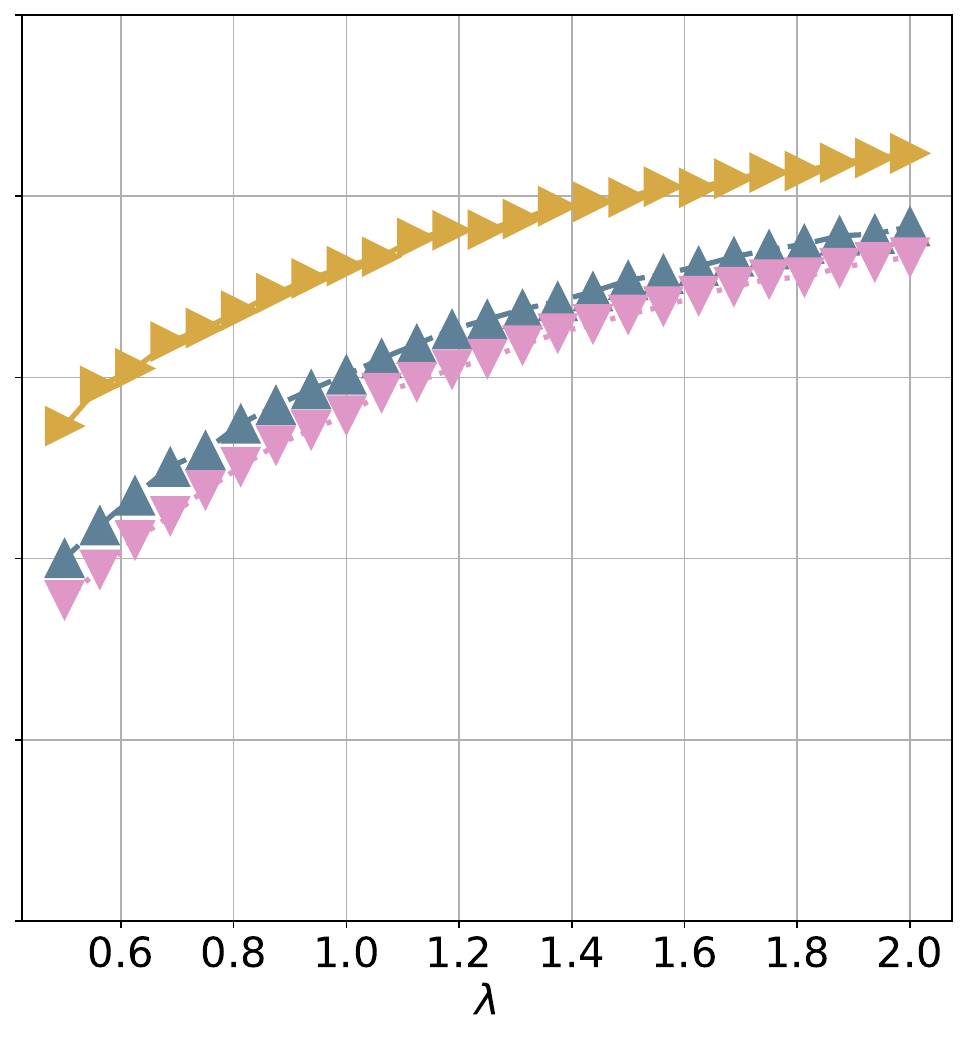}}\label{fig: component rt-twitter-copen}
\hspace{-2mm}
\caption{\textbf{The impact of edge switching on the maximum component ratio in real networks.} We set $\alpha\in\{2.60,3.70,4.60\}$ and $\lambda\in[0.5, 2.0]$ in six real-world network data sets for cross simulations. Each data point is the average of the ratios of the largest component over $t\in[80, 200]$. (a) The yeast protein network with \cite{jeong2001lethality}, (b) The inferred network by high-throughput protein-protein interactions \cite{cho2014wormnet}, (c) The inferred network by small/medium-scale protein-protein interactions \cite{cho2014wormnet}, (d) The bipartite network contains persons who appeared in at least one crime case \cite{nr-aaai15}, (e) The network of fiber tracts in brains \cite{bigbrain}, (f) The retweet and mentions network from the UN conference held in Copenhagen \cite{ahmed2010time}. }\label{fig: component in real networks}
\end{figure*}

To illustrate conclusions, we show the network topology properties in Fig.~\ref{fig: network topology}. We set the dormant time following exponential distribution $f(t)=\lambda e^{-\lambda t}, \lambda>0$, and the activated time following power-law distribution $g(t)=(\alpha-1)t^{-\alpha}, \alpha\geq 2$. Therefore, the activation constant $q_0=(\lambda\alpha-2\lambda)/(\lambda\alpha+\alpha-\lambda-1)$. We start with a random regular graph (RRG) as the underlying network $\mathcal{G}$, in Fig.~\ref{fig: network topology}a, we find that the activated edge number ($\sum_{i,j\in\mathcal{G}}x_{ij}(t)$) is stable after $t=10$, and oscillates around specific values. It is important to emphasize that this stationary state is irrelevant to the initial state because of the Markov chain irreducibility of the activated edge number. Fig.~\ref{fig: network topology}b further presents the stationary distribution of edge numbers in the mentioned stable state. $\sum_{i,j\in\mathcal{G}}x_{ij}(t)$ follows a homogeneous distribution and can be well approximated by the binomial form as Eq.~\ref{eq: binomial} suggests. The curve in the corresponding color of each parameter setting is the theoretical solution. 

Next, we select Barab{\'a}si-Albert scale-free network (BAN) \cite{barabasi1999emergence} and Watts-Strogatz small-world network (WSN) \cite{watts1998collective} as the underlying topologies. The degree distribution is a crucial property in understanding network topologies. In Fig.~\ref{fig: network topology}c, we show the degree distribution of activated subgraph in both BAN and WSN. The degree distribution is also stable with the network evolution of switching topologies. Solid curves present the theoretical result of the degree distribution in Eq.~\ref{eq: degree distribution}. An interesting phenomenon is that in BAN, the switching topology causes the collapse of a heavy-tailed degree distribution. The minimum degree in BAN is $k/2$, and we can find these vertices with the highest probability in the whole network. According to Eq.~\ref{eq: degree distribution}, these vertices have smaller degrees than $k/2$ in the activated subgraph $\mathcal{G}_A(t)$ that are controlled by the binomial term. This homogeneous disturbance leads to the peak of the degree distribution with switching topologies for BANs. 

In Figs. \ref{fig: network topology}d and e, we further present the results for the density and spectral radius of $\mathcal{G}_A(t)$. As indicated by our previous analysis, the monotonic relationship between $\lambda$ and $q_0$ is confirmed. As $\lambda$ increases, both the density and spectral radius also increase, with more pronounced growth observed for vertices with higher degrees $k$. The simulation results for density closely approximate the theoretical expression $2q_0M/N(N-1)$ discussed earlier. The activated subgraph $\mathcal{G}_A(t)$ can be interpreted as the result of random failures or attacks on the edges of the underlying network. Additionally, the underlying networks considered here are sparse. Due to the dynamic topology of the edges, bridge edges in the network can become inactive, leading to the disconnection of network components. We quantify the impact of this dynamic topology on network connectivity by measuring the relative size of the largest component in $\mathcal{G}_A(t)$, defined as the number of vertices in the largest component relative to $N$. In Fig.~\ref{fig: network topology}f, we observe that the size of the largest component is nearly identical to that of the underlying network when $k=8$. However, when the degree is halved to $k=4$, the relative size of the largest component gradually decreases as activation weakens, though it still encompasses the majority of vertices. For $k=2$, network connectivity breaks down. In the case of a BAN, the relative size of the largest component does not exceed 0.60, and in most cases, it is less than half. For a WSN, the relative size is less than 0.08, leading to a significant number of isolated components or vertices.

In addition to generated networks, we also examine the effects of edge switching on the relative sizes of the largest component in real-world network datasets. In Fig.~\ref{fig: component in real networks}, we present the maximum component ratios for six real networks with different structures. The results indicate that the largest component ratio exhibits a positive correlation with $\lambda$, but a negative correlation with $\alpha$. As $\alpha$ increases, the expected activation time approaches 1, whereas there are no such constraints on the expected dormant time as $\lambda$ increases. Notably, the curves for $\alpha = 3.70$ and $\alpha = 4.80$ are remarkably similar, suggesting that the largest component ratios are almost exclusively dependent on $\lambda$. Each network becomes fragmented when the edges exhibit a low activation tendency (i.e., when $q_0$ is small), resulting in a significant loss of robustness. 

\begin{table}[]
\centering
\begin{tabular}{cl|ccc|ccc}
\hline
\hline
\multicolumn{2}{c|}{Networks} & \multicolumn{3}{c|}{BAN}                                     & \multicolumn{3}{c}{WSN}                                      \\ \hline
\multicolumn{2}{c|}{$\left\langle k\right\rangle$}        & \multicolumn{1}{c|}{2}    & \multicolumn{1}{c|}{4}    & 8    & \multicolumn{1}{c|}{2}    & \multicolumn{1}{c|}{4}    & 8    \\ \hline
\multicolumn{2}{c|}{$\lambda=0.5$}      & \multicolumn{1}{c|}{0.22} & \multicolumn{1}{c|}{0.85} & 0.98 & \multicolumn{1}{c|}{0.02} & \multicolumn{1}{c|}{0.88} & 0.99 \\ \hline
\multicolumn{2}{c|}{$\lambda=1.5$}      & \multicolumn{1}{c|}{0.32} & \multicolumn{1}{c|}{0.97} & 0.99 & \multicolumn{1}{c|}{0.05} & \multicolumn{1}{c|}{0.99} & 0.99 \\ \hline
\multicolumn{2}{c|}{$\lambda=2$}        & \multicolumn{1}{c|}{0.52} & \multicolumn{1}{c|}{0.98} & 0.99 & \multicolumn{1}{c|}{0.06} & \multicolumn{1}{c|}{0.99} & 0.99 \\ \hline \hline
\end{tabular}
\caption{\textbf{The largest component ratio in synthetic networks given $\alpha=2.60$.}}\label{tab: 1}
\end{table}

\begin{table}[]
\centering
\begin{tabular}{cl|c|c|c|c|c|c}
\hline\hline
\multicolumn{2}{c|}{Networks} & Yeast & WormNet-CE-HT & WormNet-CE-LC & Moreno Crime & Brain & Retweet \\ \hline
\multicolumn{2}{c|}{$\left\langle k\right\rangle$}        & 2.67  & 2.28          & 2.37          & 3.56         & 3.03  & 2.70    \\ \hline
\multicolumn{2}{c|}{$\lambda=0.5$}      & 0.53  & 0.38          & 0.37          & 0.76         & 0.65  & 0.55    \\ \hline
\multicolumn{2}{c|}{$\lambda=1.5$}      & 0.79  & 0.67          & 0.54          & 0.92         & 0.82  & 0.79    \\ \hline
\multicolumn{2}{c|}{$\lambda=2$}        & 0.85  & 0.72          & 0.60          & 0.95         & 0.86  & 0.85    \\ \hline\hline 
\end{tabular}
\caption{\textbf{The largest component ratio in real network data sets given $\alpha=2.60$.}}\label{tab: 2}
\end{table}

In Tabs. \ref{tab: 1} and \ref{tab: 2}, we further summarize the exact largest component ratio for synthetic and real networks respectively with $\alpha=2.60$. As the mean degree increases, the largest component nearly occupies the whole network. For a network with a small mean degree ($\left\langle k\right\rangle<4$), the components can be mostly separated, and the largest component ratio is small as well even if the activation is strong (e.g., $\lambda=1.5$). In real networks that we sample, the mean degrees are often small especially for biological networks. Therefore, the largest component ratio grows slowly as the increase of $\lambda$. 

Based on the results above, we find that the switching topology can lead to the collapse of the network structure. Next, we study the influence of the switching topology on two dynamic processes (random walks and evolutionary dynamics). 
\subsection{Deceleration of Random Walks}

\begin{figure*}[h]                               
\centering
\subfigure[Random Walk Example]{\includegraphics[scale=0.07]{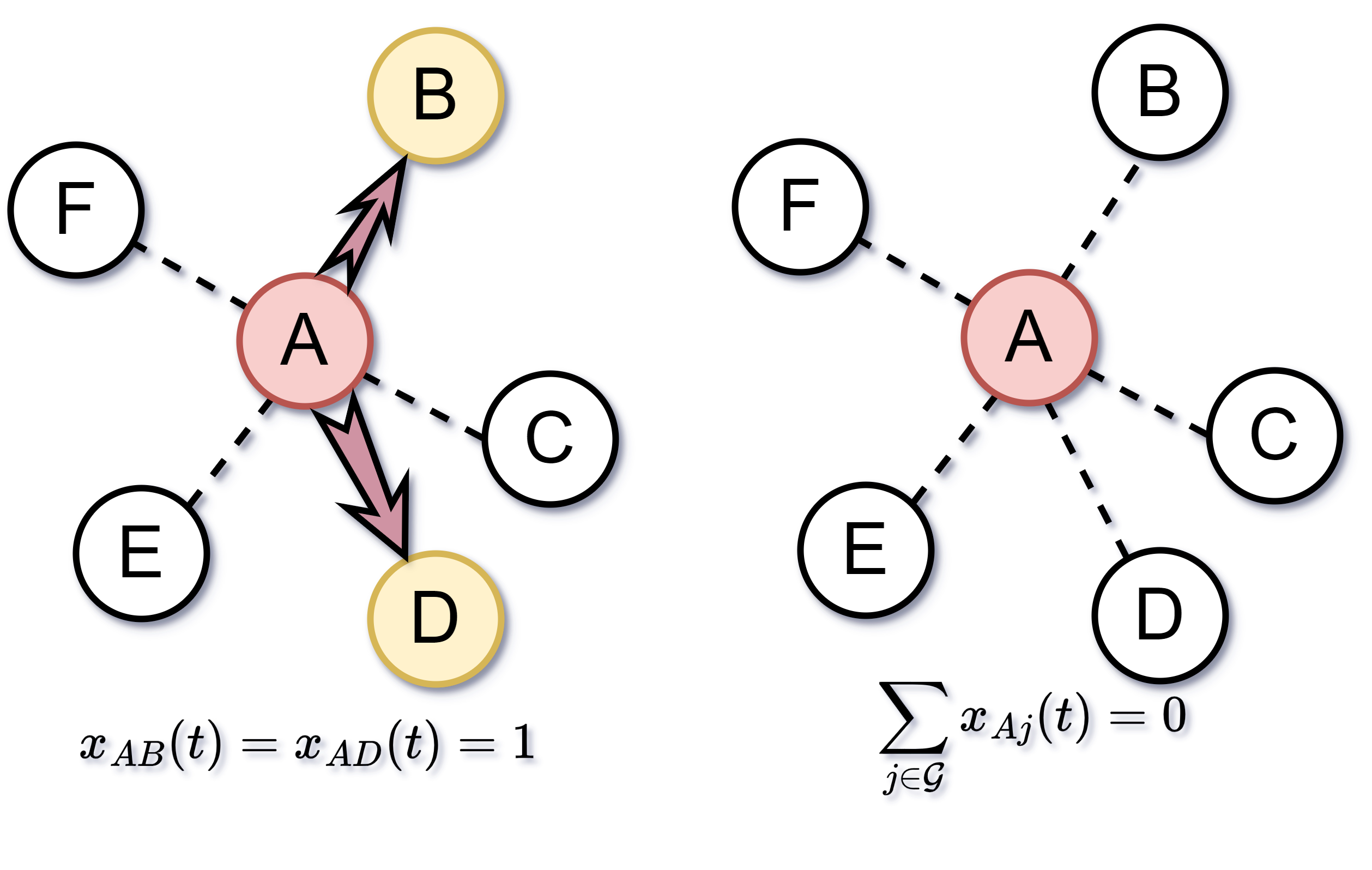}}\label{fig: random walk example}
\vspace{-2mm}
\subfigure[RRG]{\includegraphics[scale=0.25]{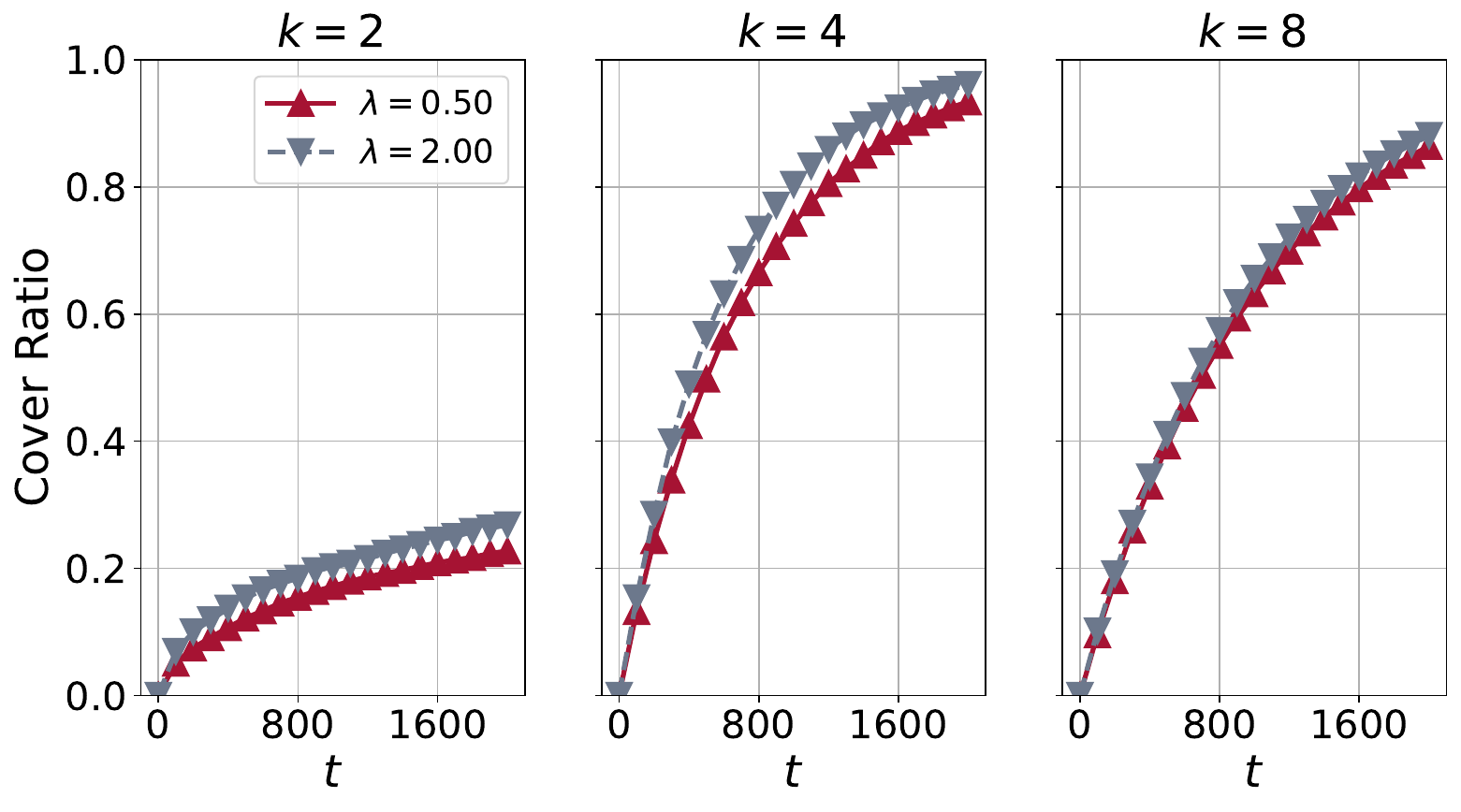}}\label{fig: coverRRG}
\vspace{-2mm}

\subfigure[WSN]{\includegraphics[scale=0.24]{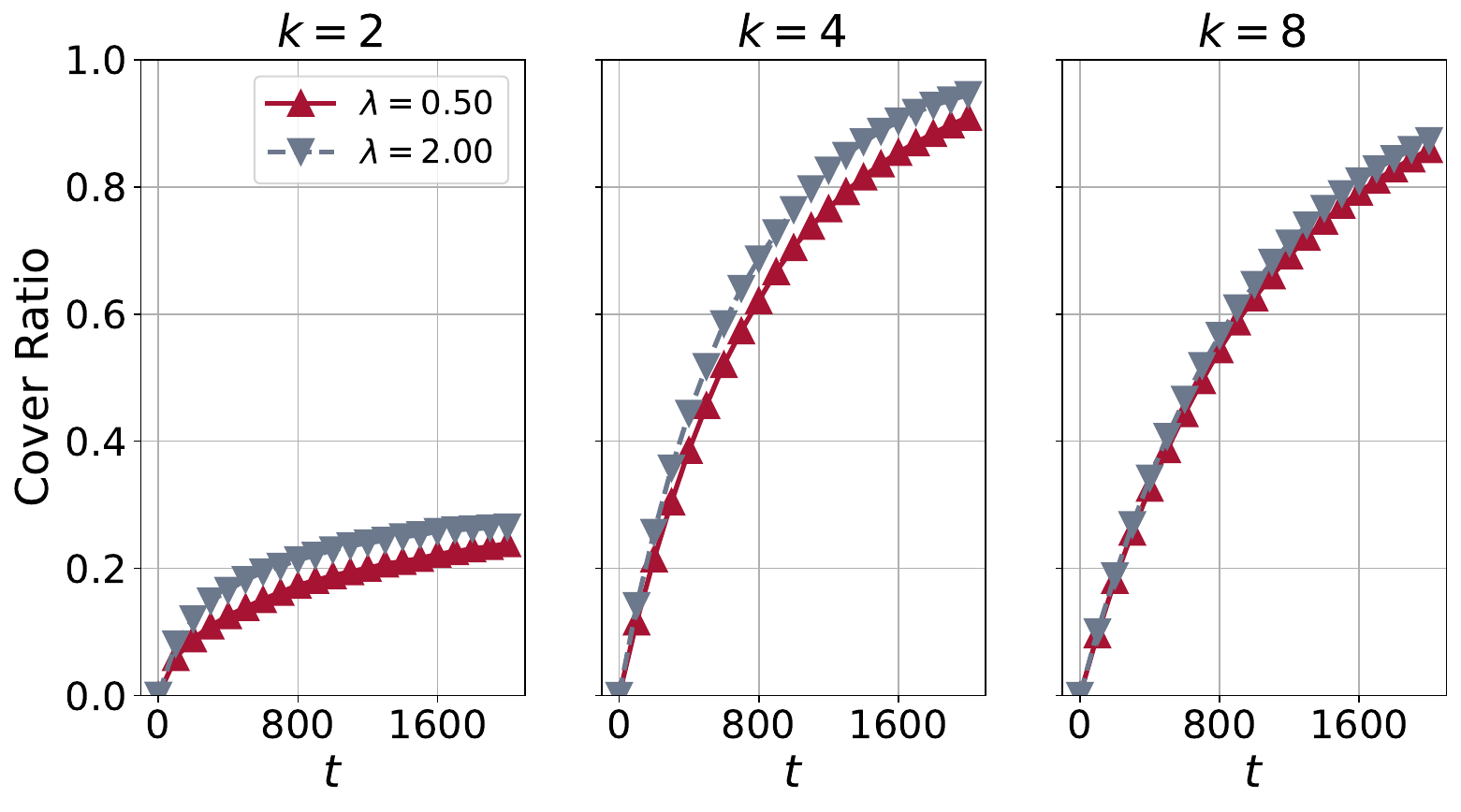}}\label{fig: coverWSN}
\hspace{-2mm}
\subfigure[BAN]{\includegraphics[scale=0.24]{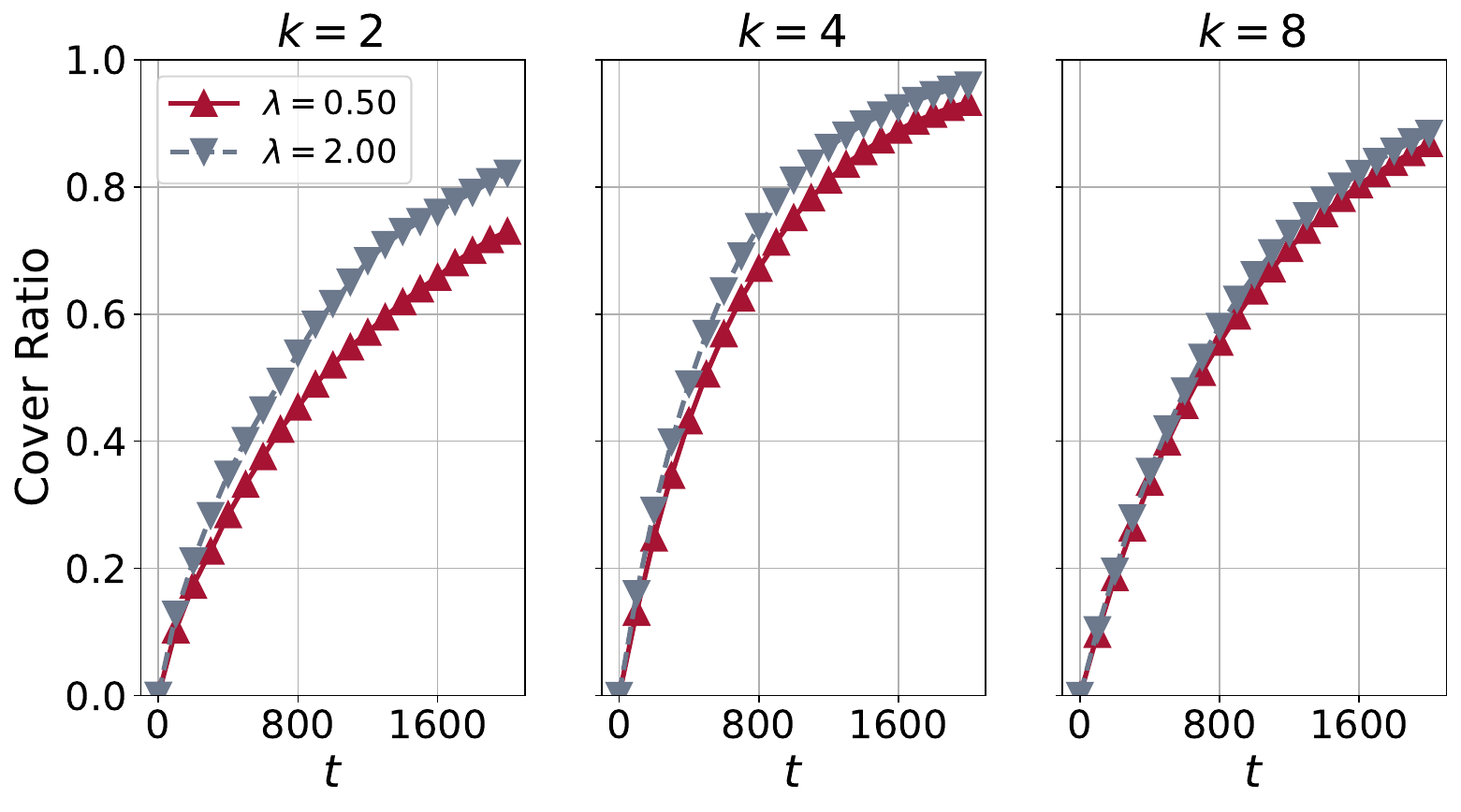}}\label{fig: coverBAN}
\vspace{-2mm}
\caption{\textbf{Edge cover ratio for random walks. } (a) An example of random walks with network switching topology. Our focal random walk steps by Poisson process and is now in vertex $A$. In the left panel, edges $(A, B)$ and $(A, D)$ are activated, and the focal walker can step into $B$ or $D$ with equal probability. In the right panel, all edges around $A$ are dormant, and the walker remains in $A$. (b)-(d) present the edge cover ratio against time in RRG, WSN, and BAN respectively, with $N=200$, $\lambda\in\{0.50, 2.00\}$, $\alpha=2.60$, and $k\in\{2,4,8\}$. Each data point is obtained by the average of 100 independent realizations. }\label{fig: random walk cover}
\end{figure*}

A random walk describes the simple information dynamics in a complex network \cite{hoffmann2012generalized, speidel2015steady}. In the context of switching topology, we assume that a walker ${R(t) \in \mathcal{V}, t \geq 0}$ starts from a random vertex and moves to the next position according to a Poisson process with rate parameter 1. The walker continuously steps to one of its neighboring vertices through activated edges. If no activated edge is available, the walker remains at its current position. The walker can move from vertex $i$ to vertex $j$ only if $x_{ij}(t) = 1$. An example illustrating when the walker can or cannot move to neighboring vertices is shown in Fig.~\ref{fig: random walk cover}a. In the left panel, we observe that $x_{AB}(t) = x_{AD}(t) = 1$, meaning the walker at vertex $A$ can randomly step to either neighboring vertex $B$ or $D$ via the activated edges. In contrast, the right panel shows a situation where no neighboring edges are activated, causing the walker to remain at vertex $A$ during that round.

We can discretize this random walk process into a discrete-time Markov chain that records only the walker’s transitions. Let $l_{ij}$ denote the probability that the walker $R(t)$ was at vertex $i$ in the previous transition and will next move to vertex $j$. Since the probability of an edge being activated is $q_0$, we can express the transition probability as (see Methods \ref{Methods: 2} for details):
\begin{equation}\label{eq: one step random walk}
l_{ij}=\left\{
\begin{aligned}
&\frac{1-(1-q_0)^{k_i}}{k_i}  &,  (i,j)\in \mathcal{E}\\
&(1-q_0)^{k_i}&,  i=j\\
&0&,  other
\end{aligned}
\right.
\end{equation}
where $k_i$ is vertex $i$'s degree in the underlying network $\mathcal{G}$. In fact, $l_{ij}$ indicates the probability for a one-step random walk from vertex $i$ to $j$. Since the random walk $R(t)$ is an irreducible stochastic process for a strongly connected underlying network, i.e., each state is a recurrent state. After sufficient time $t$, we have the probability of finding the walker in vertex $i$ (see Methods \ref{Methods: 2} for details)
\begin{equation}\label{eq: random walk stationary}
P(R(t)=i)=\frac{k_i}{[1-(1-q_0)^{k_i}] \sum_{j\in\mathcal{G}}\frac{k_j}{1-(1-q_0)^{k_j}}}. 
\end{equation}
For isothermal underlying networks, this stationary distribution degenerates to $P(R(t)=i)=1/N$. Therefore, the random walk in an isothermal graph with the switching topology has the same probability of dwelling in each vertex. 

We assume that when a walker steps from vertex $i$ to vertex $j$, the corresponding edge $(i,j)$ is considered covered by the random walk. If the underlying network is strongly connected, the walker will eventually traverse all edges given sufficient time. Our focus is on the edge cover rate of a random walk in $\mathcal{G}_A(t)$, specifically how the number of covered edges evolves over time. To investigate this, we select three underlying networks: Random Regular Graph (RRG), Watts-Strogatz Network (WSN), and Barabási-Albert Network (BAN), and numerically study the edge cover rate under different network configurations.

From the results in Figs. \ref{fig: random walk cover}b–d, we observe that when $k=2$, the edge cover rates of RRG and WSN are significantly slower than that of BAN. For a vertex with degree 2, the probability of becoming isolated is $(1-q_0)^2$. If a random walker encounters such an isolated vertex, the likelihood of the walker being trapped is relatively high. As mentioned earlier, a snapshot of the activated subgraph $\mathcal{G}_A(t)$ can be viewed as representing random failures or attacks on edges in the underlying network. BAN, with its hub vertices, is robust against such random failures and thus demonstrates resilience to temporary structural deficiencies caused by the stochastic switching of the topology.

We have previously discussed the impact of switching topology on connected components. A random walker is also susceptible to being trapped in an isolated component, which affects the overall edge cover rate. As shown in Figs. \ref{fig: random walk cover}b–d, with a higher $\lambda$ (i.e., a higher $q_0$), the edge cover rate of the random walk increases in networks with lower degrees. This enhancement is particularly pronounced in Barabási-Albert Networks (BAN). However, as the degree of the underlying network increases, the influence of $\lambda$ diminishes. When comparing the edge cover rates for $k=2$ and $k=4$, we observe a substantial increase for $k=4$, but a decrease in cover speed when $k=8$. This phenomenon can be explained by the probability of a single vertex becoming trapped. For instance, at $\lambda = 0.50$, the trapped probability is approximately $(1 - q_0)^2 \approx 0.184$ for $k=2$, but it drops significantly to $(1 - q_0)^4 \approx 0.034$ for $k=4$ and $(1 - q_0)^8 \approx 0.001$ for $k=8$. In the latter two cases, the probability of a random walker being trapped at a single vertex or within a connected component is similarly low. However, the number of edges that the walker must traverse is twice as large for $k=8$ compared to $k=4$. As a result, the edge cover rate for $k=8$ grows more slowly than for $k=4$.

\begin{figure*}[h]                                 
\centering
\subfigure[Yeast]{\includegraphics[scale=0.24]{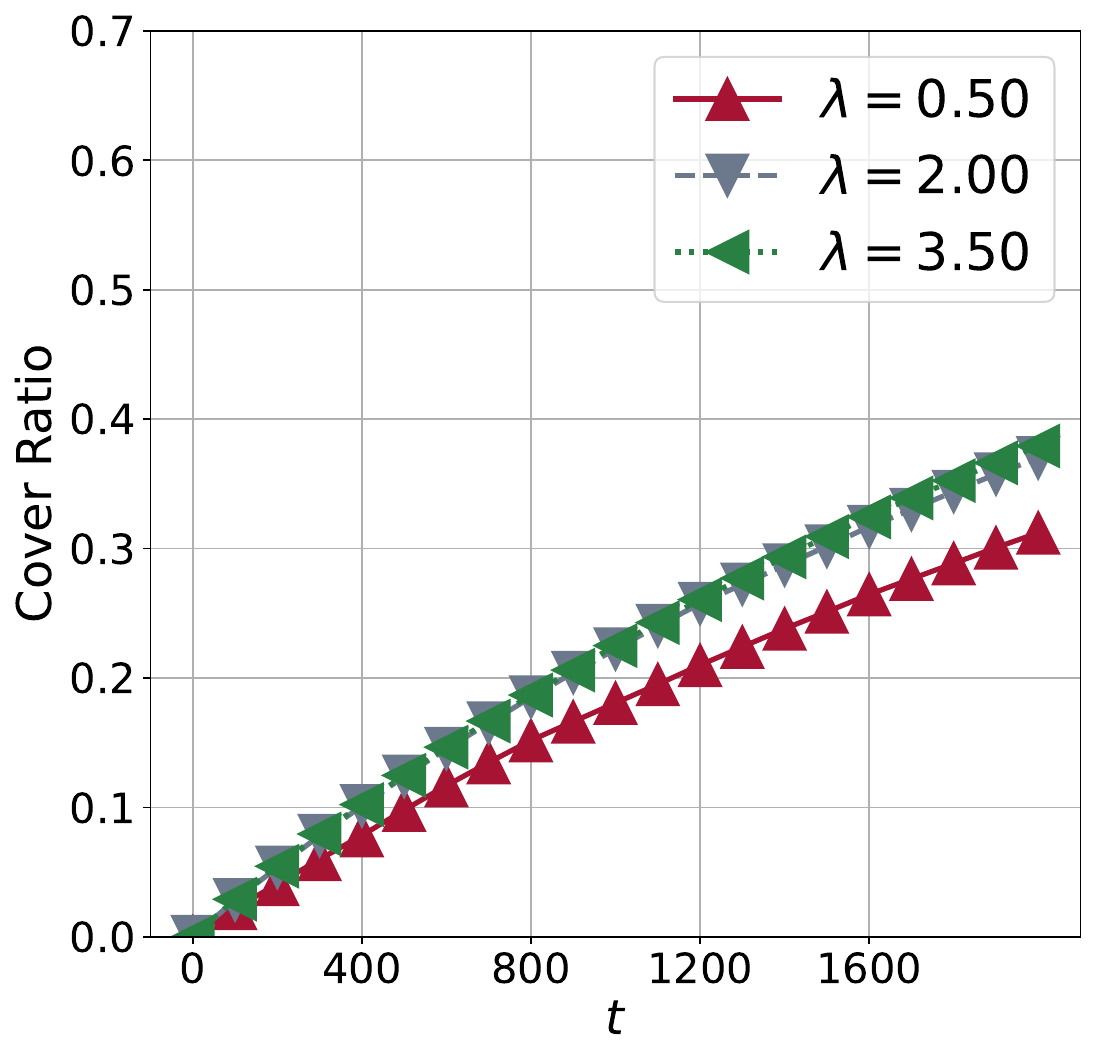}}\label{fig: cover bio-yeast}
\hspace{-2mm}
\subfigure[WormNet-CE-HT]{\includegraphics[scale=0.24]{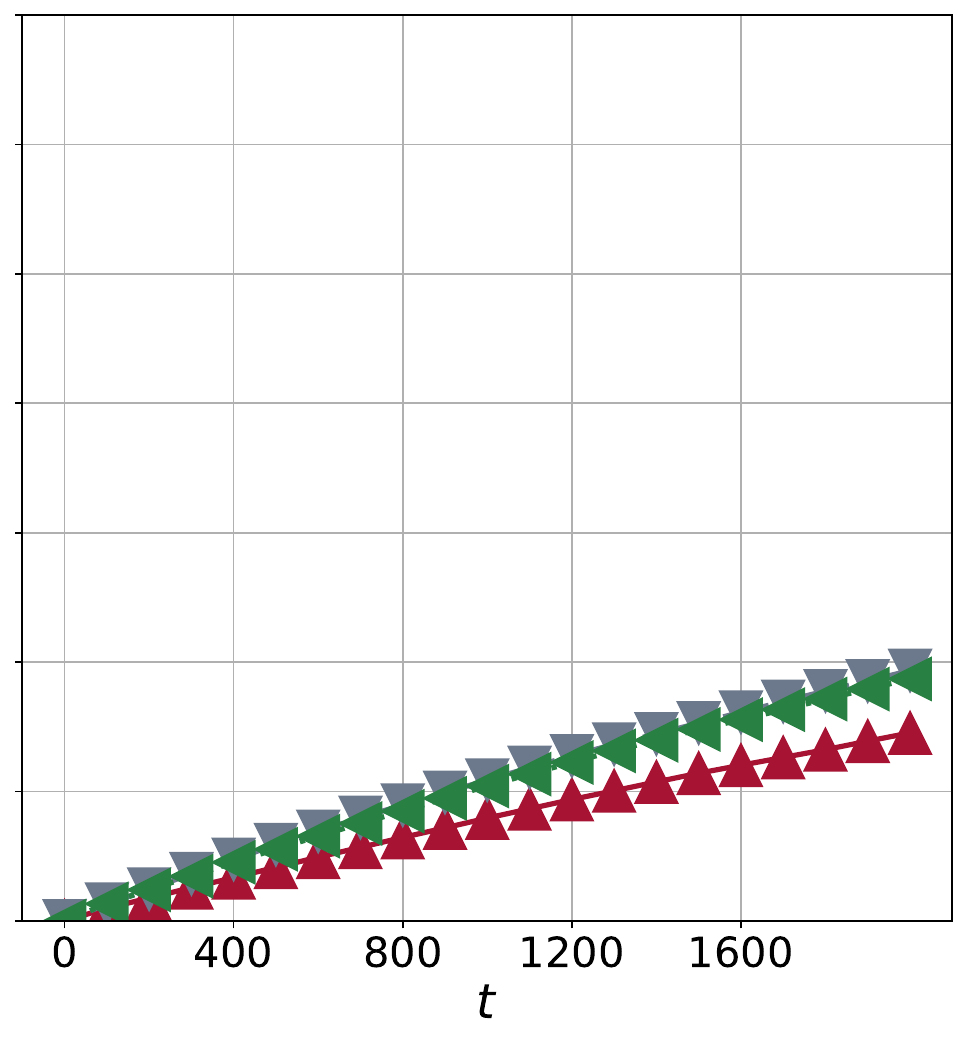}}\label{fig: cover bio-CE-HT}
\hspace{-2mm}
\subfigure[WormNet-CE-LC]{\includegraphics[scale=0.24]{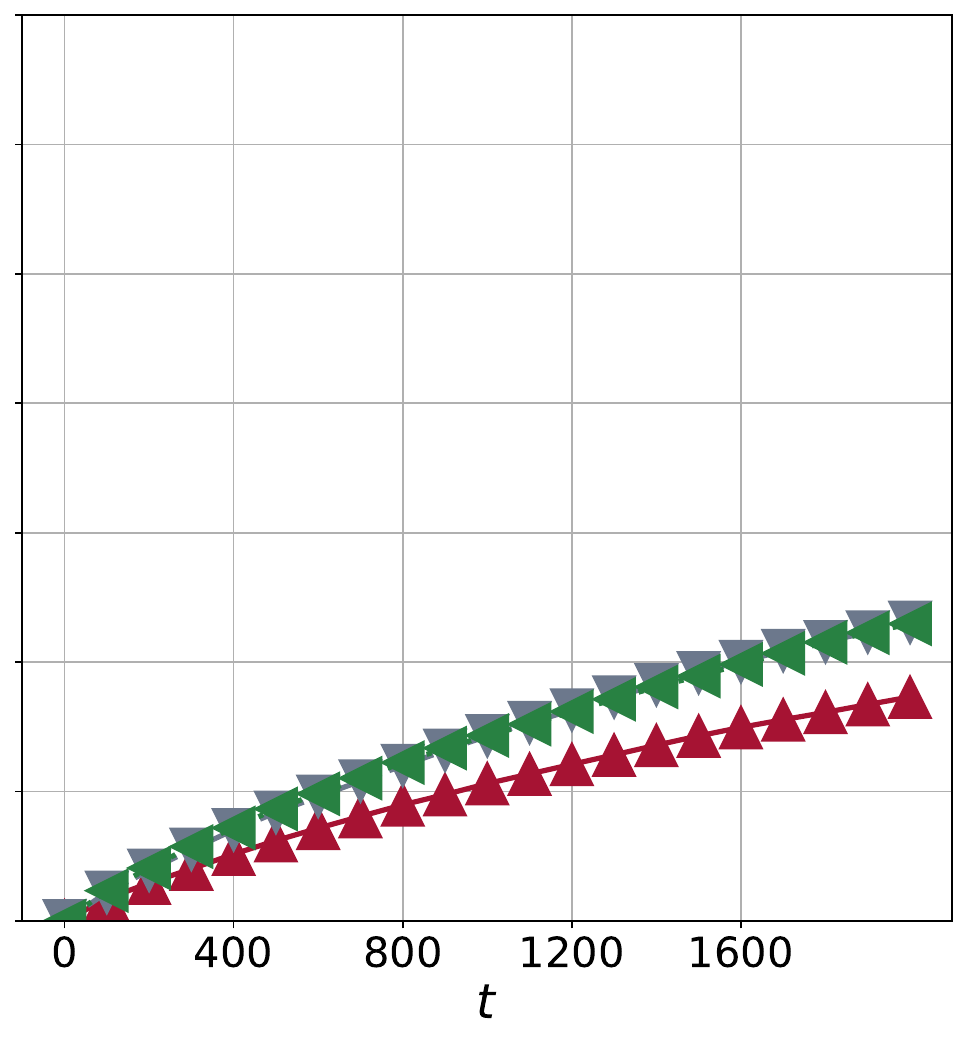}}\label{fig: cover bio-CE-LC}
\hspace{-2mm}

\subfigure[Moreno Crime]{\includegraphics[scale=0.24]{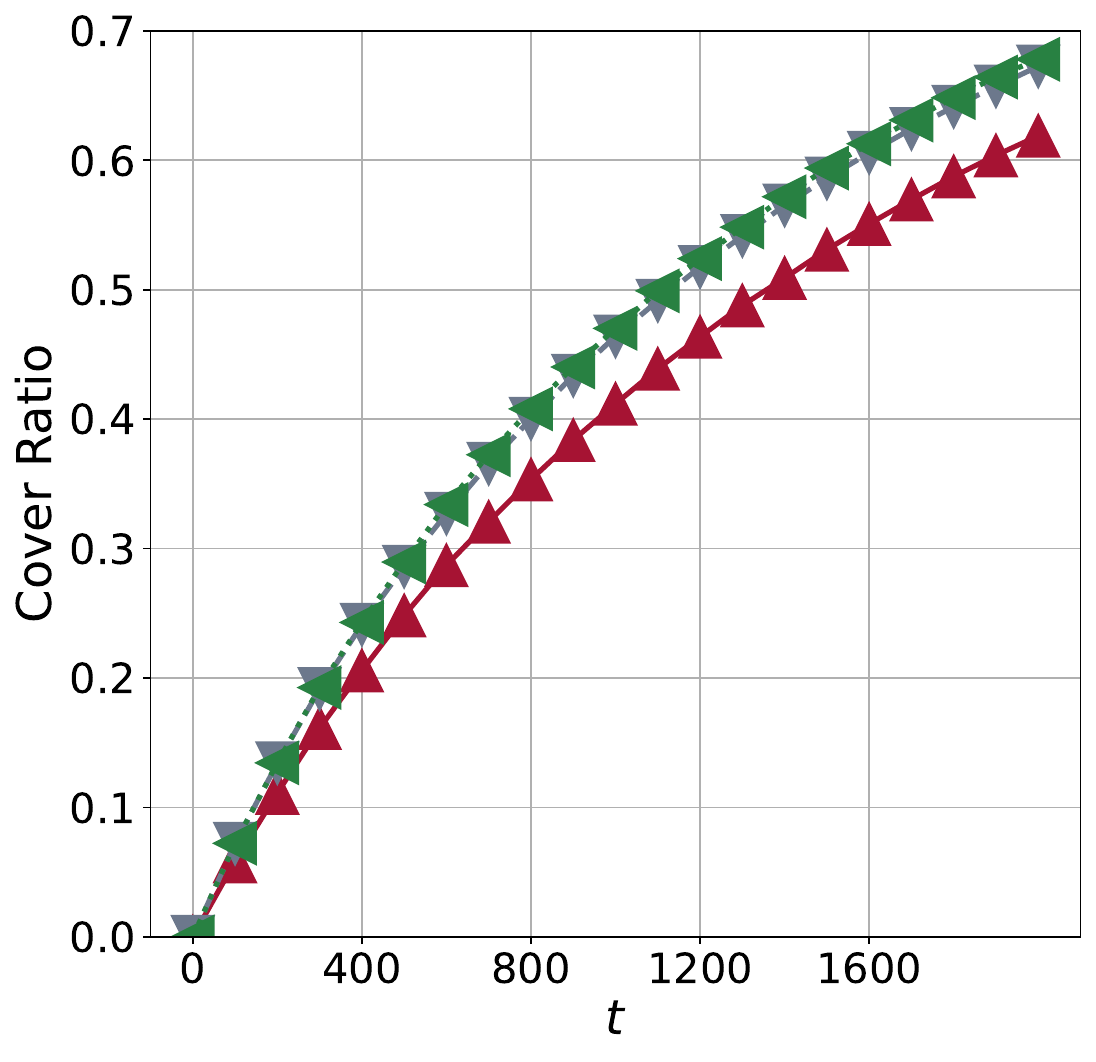}}\label{fig: cover ia-crime-moreno}
\hspace{-2mm}
\subfigure[Brain]{\includegraphics[scale=0.24]{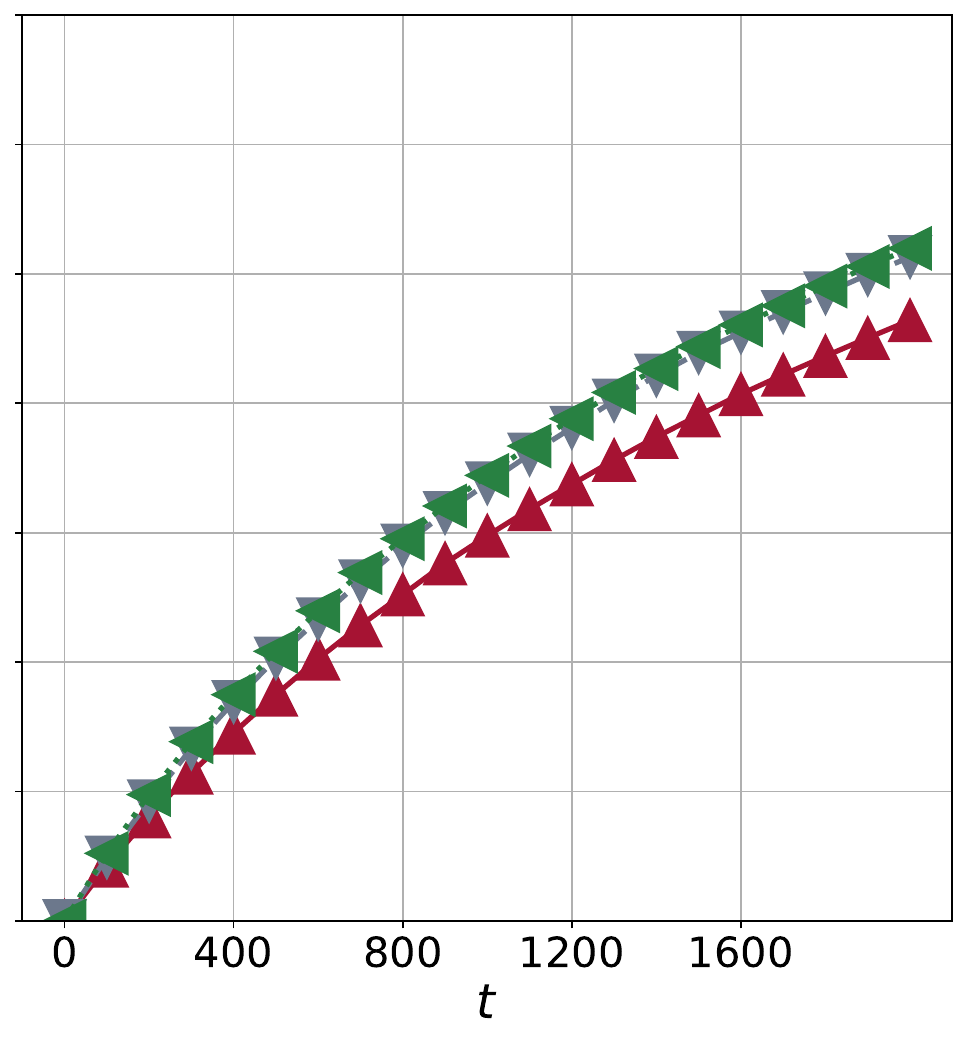}}\label{fig: cover bn-mouse-kasthuri_graph_v4}
\hspace{-2mm}
\subfigure[Retweet]{\includegraphics[scale=0.24]{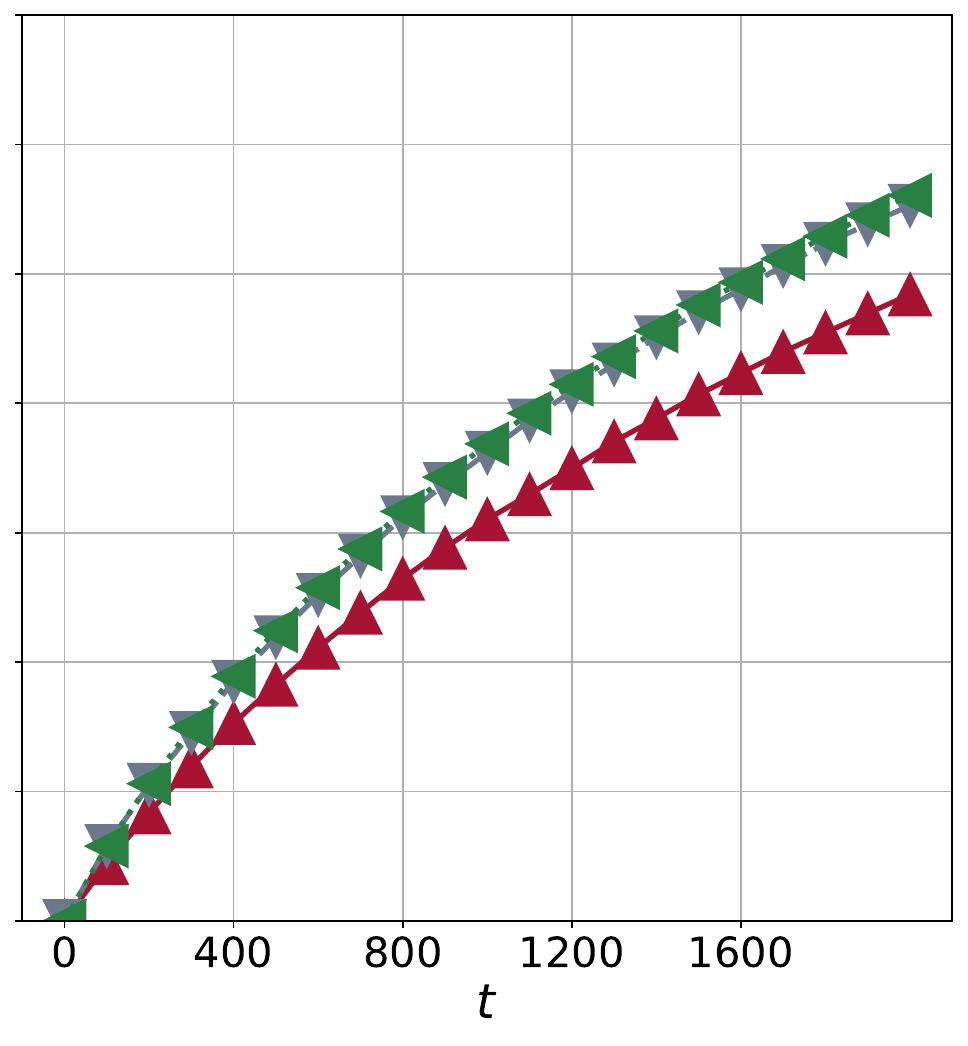}}\label{fig: cover rt-twitter-copen}
\hspace{-2mm}
\caption{\textbf{Vertex cover rates of random walks in real networks. }We set $\lambda\in\{0.50, 2.00, 3.50\}$ and $\alpha=2.60$ for cross simulations. We observe the random walks during $t\in[0,2000]$. Each data point is obtained by the average of 100 independent runs with randomly selected initial vertex. The settings of real networks are the same as Fig.~\ref{fig: component in real networks}. }\label{fig: random walk in real networks}
\end{figure*}

In Fig.~\ref{fig: random walk in real networks}, we further analyze the vertex cover ratio in the real-world networks previously discussed in Fig.~\ref{fig: component in real networks}. Our findings indicate that $\lambda = 0.50$ results in the slowest vertex cover rate across all networks compared to the other two cases. At this setting, $q_0 = 0.125$, which leads to a high likelihood of vertices, particularly those with small degrees, becoming isolated. As a result, the walker is more likely to be trapped at these vertices or within their associated components. However, the results for $\lambda = 2.00$ and $\lambda = 3.50$ exhibit less clear trends across the networks. For instance, in Fig.~\ref{fig: random walk in real networks}a, $\lambda = 3.50$ generally yields a slightly higher vertex cover ratio, while in other cases, such as Fig.~\ref{fig: random walk in real networks}b and c, $\lambda = 2.00$ tends to result in a higher cover rate.

\subsection{Promotion of Cooperation}
Evolutionary dynamics is a powerful framework for studying cooperation \cite{ling2025supervised,mohamadichamgavi2024impact,yue2025coevolution} and behavioral economics \cite{10289641} in populations. Since Darwin, researchers have sought to uncover the mechanisms behind evolution. In the prisoner's dilemma game, spatial structure allows a population to overcome the Nash equilibrium of defection, a phenomenon known as network reciprocity. Recently, the evolution of cooperation in time-varying networks has garnered significant attention\cite{li2020evolution, su2023strategy}.

In this study, we employ the prisoner's dilemma (PD) model to explore the effects of switching topology on evolutionary dynamics. In this model, two types of individuals—cooperators (C) and defectors (D)—compete within the population. Each individual occupies a vertex, and edges in the underlying network represent potential interactions between individuals. Mutual cooperation yields a reward of $b - c$ for both cooperators, while mutual defection results in a punishment of $0$ for both defectors. In the case of one-way cooperation, the cooperator receives the sucker's payoff of $-c$, while the defector gains the temptation payoff of $b$. The ratio $b/c$ is referred to as the benefit-to-cost ratio. The payoff for individual $i$ ($\pi_i$) is the sum of rewards from all activated connections with neighbors. The fecundity of an individual is modeled by an exponential function of the payoff, $F_i = \exp(w \pi_i)$, where $0 < w \ll 1$ represents the strength of weak selection (see Methods \ref{Methods: 3} for details).

\begin{figure*}                                 
\centering
\subfigure[Fecundity Calculation and Replacement]{\includegraphics[scale=0.15]{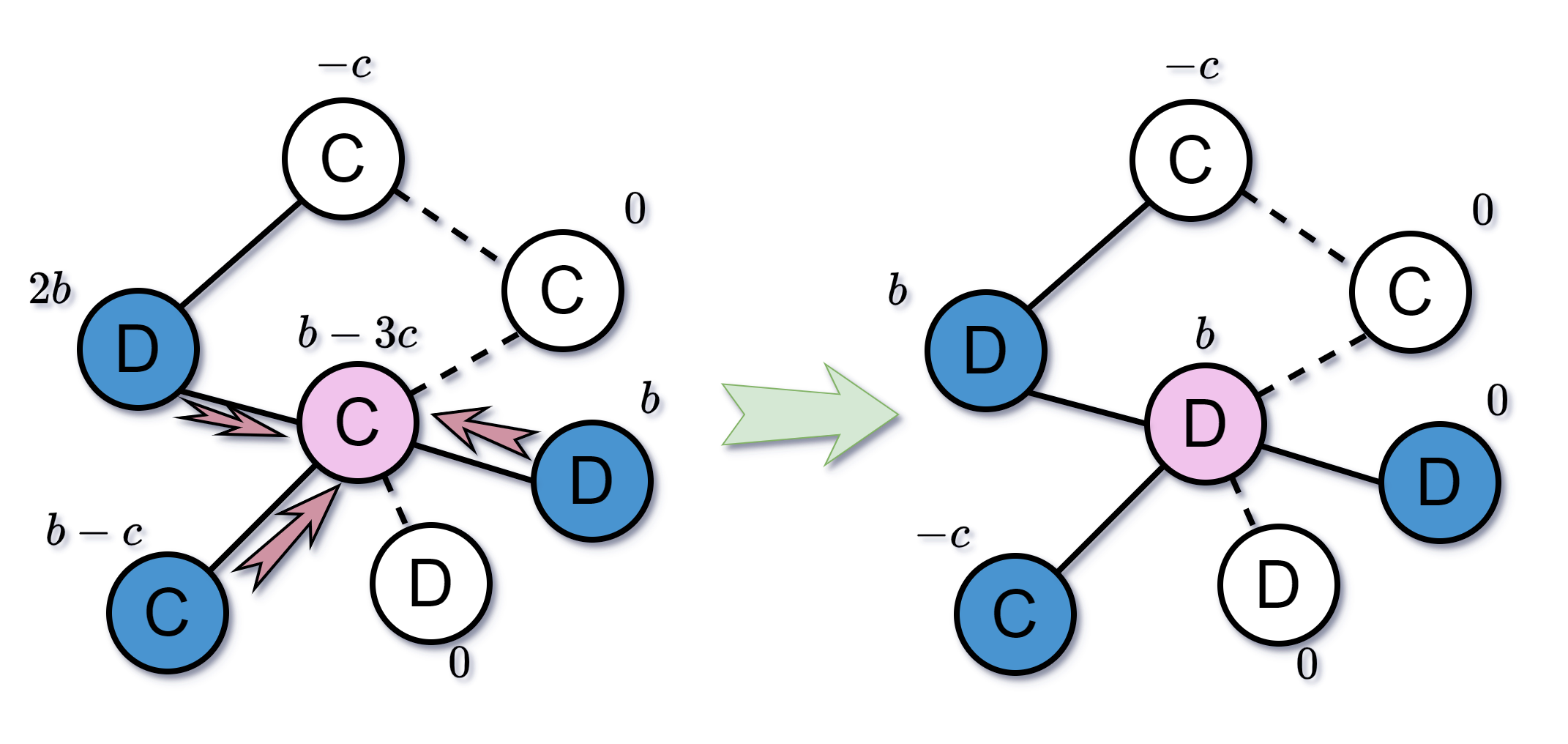}}\label{fig: game example}

\subfigure[RRG, $k=4$]{\includegraphics[scale=0.24]{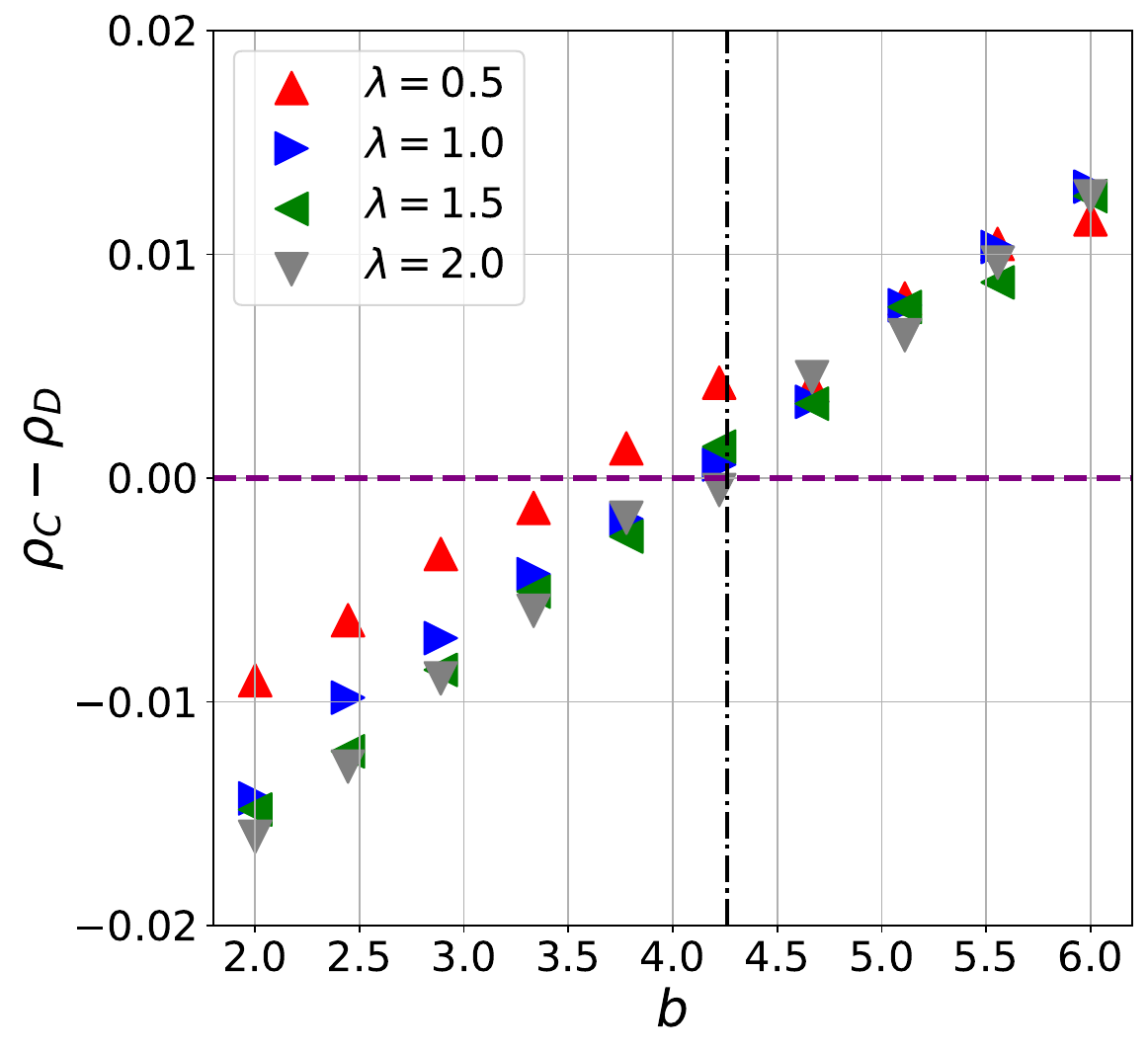}}\label{fig: RRGGame4}
\subfigure[RRG, $k=6$]{\includegraphics[scale=0.24]{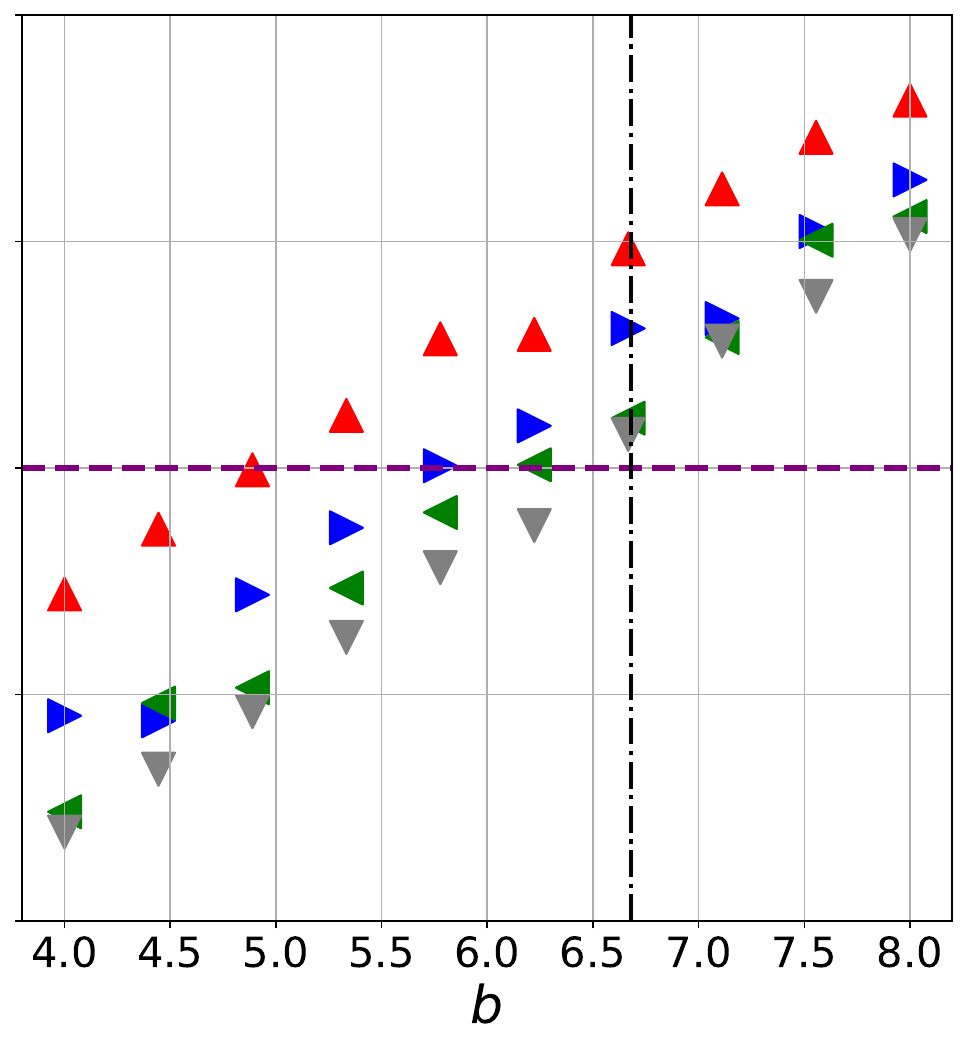}}\label{fig: RRGGame6}
\subfigure[RRG, $k=8$]{\includegraphics[scale=0.24]{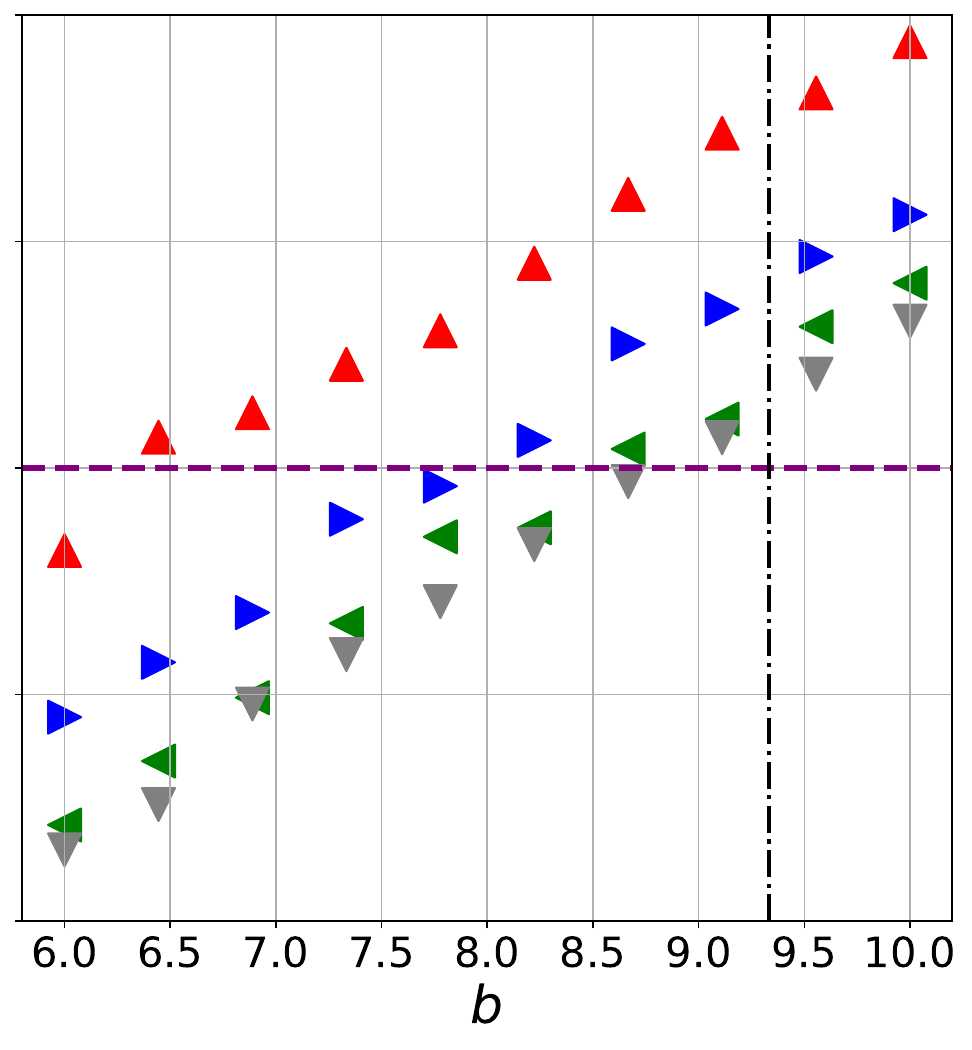}}\label{fig: RRGGame8}
\caption{\textbf{Fixation probability of evolutionary dynamics. } (a) An example of the evolutionary donation game with the network switching topology. The fecundity calculation and replacement event from neighbors occur only by activated edges around the focal updating vertex. In this example, our focal cooperator in the center has three activated neighbors and obtains $b-3c$ as the payoff. Through the death-birth process, it turns into a defector. (b)-(d) $\rho_C-\rho_D$ against temptation $b$ in RRGs. The settings are $c=1$, $N=100$, $k\in\{4,6,8\}$, and $\lambda\in\{0.5,1.0,1.5,2.0\}$. The vertical dashed lines denote the baselines $(N-2)/(N/k-2)$ for $\rho_C>\rho_D$ without the switching topology. The purple dashed line denotes the neutral drift, where $\rho_C=\rho_D$. $\rho_C$ and $\rho_D$ are obtained by computing the ratio of fixation times in $5\times10^4$ independent experiments respectively. (b) $k=4$. (c) $k=6$. (d) $k=8$. }\label{fig: evolution game}
\end{figure*}

Under the switching topology, each participant's payoff is determined by interactions with neighbors through activated edges. Given the continuous-time framework, the replacement process follows a death-birth mechanism: individuals are selected to die according to independent Poisson processes, after which their neighbors compete for the vacant position with probabilities proportional to their fecundity through activated connections. If no activated edge exists around the focal individual, the individual retains its current strategy. An example of this strategy replacement process is illustrated in Fig.~\ref{fig: evolution game}a.

The fixation probability of cooperation, $\rho_C$, (respectively, defection, $\rho_D$) is defined as the probability that a single cooperator (respectively, defector) invades the network and eventually dominates the population. Cooperation is favored when $\rho_C > \rho_D$. To provide a comparison with the condition for cooperation in static networks, we use Random Regular Graphs (RRG) and several real-world networks to minimize the uncertainties associated with random network structures. In a static network, the condition for cooperation is given by $(N-2)/(N/k-2)$, where $N$ is the population size and $k$ is the degree of the network.

Our results in Figs. \ref{fig: evolution game}b–d indicate that the switching topology lowers the threshold for $\rho_C - \rho_D > 0$, thereby promoting the evolution of cooperation. This promotion is particularly pronounced for smaller values of $\lambda$ and in networks with larger degrees. It is important to note that the monotonic relationship between $\lambda$ and $q_0$ implies that reducing edge activation enhances the diffusion of cooperative behavior. In networks without switching topology, the condition for cooperation typically correlates positively with network degree, which can impose strict constraints on the evolution of cooperation in regular networks. The proposed switching topology, however, facilitates the emergence of cooperation at a relatively lower cost, especially in scenarios with weak activation and high network degrees. The promotion of cooperation cannot be quantified using the mean degree of the network with switching topology. There is a subtle relationship between the cooperation condition and the network topology. 
\section{Conclusions and Discussions}\label{sec3}
We propose a continuous-time switching topology model in complex networks driven by bursty behaviors, characterized by a general interevent time distribution, to describe the evolution of temporal links. This switching topology mechanism captures the intermittent interactions commonly observed in social networks. Our primary focus is on the stationary properties of the network structure under this switching mechanism, including the number of activated edges ($\sum_{i,j\in\mathcal{G}}x_{ij}(t)$) and the degree distribution of the activated subgraph $\mathcal{G}_A(t)$. The intermittent transition of edges between activated and dormant states can significantly destabilize network topologies. Based on our simulation results, we conclude that a reduction in activation decreases the number of activated edges, density, connectivity, and spectral radius of the network. Our model allows a flexible combination of activation and dormant time distributions. If other interevent time distributions are considered, the outcomes should closely follow our theorems regarding the variation of $q_0$. Accordingly, in our future work, we aim to predict the topology of social and communication networks exhibiting power-law patterns \cite{zeng2025complex}, as well as brain dynamics, which often present a mixed case \cite{wang2019non,lombardi2020critical}. 

We examine the impact of the switching topology on two dynamic processes: random walks and evolutionary game dynamics. Our connectivity results reveal that the switching topology can result in isolated components or vertices, which substantially slows down the edge coverage during random walks. When the activation constant or degree is small (i.e., edge activity is weak), random walkers are likely to be trapped within isolated components or vertices, thereby delaying the coverage and spread of information. This phenomenon can lead to a temporary "information cocoon" \cite{piao2023human, hou2023information, santos2023break} among the networked population, persisting until connections or bridges to other components are reactivated. Further investigation is needed to fully understand the conditions necessary to avoid such information cocoons caused by intermittent switching topologies.

We also explore the effect of the switching topology on the evolution of cooperation using donation game theory. Our findings suggest that the switching topology lowers the threshold for $\rho_C > \rho_D$, thereby promoting the evolution of cooperation in networked populations. This effect is particularly pronounced in networks with higher degrees and weaker activation, compared to static networks without switching topologies. From a theoretical perspective, considering incomplete information \cite{wang2023imitation} in the evolutionary process complicates the analysis, as the number of accessible neighbors varies over time according to the renewal process of each vertex. \cite{su2023strategy} successfully proposed a framework for calculating the conditions for cooperation in dynamic networks by computing the transition matrix of all potential network states. In an underlying network $\mathcal{G}$ with $M$ edges, there are $2^M$ possible activated subgraphs, leading to a $2^M \times 2^M$ transition matrix. This results in an exponentially increasing time complexity. Further research is required to precisely quantify the effect of switching topologies on evolutionary dynamics and cooperation. 

Our model can be extended by considering other assumptions to approximate the real cases. For example, the activation of one edge may influence another. Therefore, we can further study the effects of dependent activation and dormant time distributions on both the network topology and dynamics. The effect of vertex-switching on network topology can also be further explored and compared to the edge-switching. Multi-layer networks and higher-order networks can also be taken into account. More theorem on dynamic processes, e.g., specific condition for evolutionary cooperation, epidemic propagation, and synchronization, are also interesting questions regarding the bursty switching topology. 
\section{Methods}
\subsection{Stationarity of Switching Topology}\label{Methods: 1}
For one single edge, its state undergoes a cyclic but not periodic process between two states, i.e., activated and dormant. Obviously, this stochastic process is a recurrent Markov chain with two states and thus becomes renewal and regenerative, because it can be regarded as a restarting process once it turns into the initial state again. For the mentioned concept, a figurative example is that a light can be turned on and off as a cyclic process. For simplicity, we define this stochastic process undergoes a cycle once an edge state first returns to the initial state from the initial state. The limit probability of an edge state can be described by the expected duration proportion in one single cycle. We can find one edge in the activated state with the probability $\int tf(t)dt/\int t(f(t)+g(t))dt$ (defined as the activation constant $q_0$) and in the dormant state with the probability $\int tg(t)dt/\int t(f(t)+g(t))dt$ after a sufficient time $t$, which are the expected time proportion for an edge to be activated and dormant in a cycle respectively. Accordingly, the limit probability that we find $m$ activated edges in the focal network is the binomial form given in Eq.~\ref{eq: binomial}, due to our independent activation assumption. This conclusion leads to other corollaries of the activated subgraph $\mathcal{G}_A(t)$. For $\mathcal{G}_A(t)$, the expected activated edge number is $E[\vert\mathcal{E}_A\vert]=q_0 M$, the expected density is 
$E[\rho]=2q_0M/N(N-1)$, and the expected mean degree is 
$E[\left\langle k\right\rangle]=2q_0 M/N$. 

We can obtain the probability to find the focal vertex with degree $i$ in $\mathcal{G}_A(t)$ while with degree $j$ in $\mathcal{G}$ as 
\begin{equation}
p_A(i\vert j)=\frac{j!}{i!(j-i)!} q_0^i (1-q_0)^{j-i} p(j). 
\end{equation}
Note that finding $i$ neighbors of the focal vertex in $\mathcal{G}_A(t)$ if and only if its degree in $\mathcal{G}$ is greater than or equal to $i$. Sum all the terms ($p_A(i\vert j)$) with the condition $j\geq i$ leads to the expected degree distribution of $\mathcal{G}_A(t)$ as Eq.~\ref{eq: degree distribution}. 
\subsection{Random Walk Probability}\label{Methods: 2}
The stochastic process $\{R(t)\in\mathcal{V}, t\geq 0\}$ is a continuous, traversal, irreducible Markov chain. Directly solving the stationarity of random walk is difficult. We transfer this continuous-time process to a discrete-time Markov chain considering each Poisson event in the random walk as a one-step transition, inducing a new process $\{R'(t')\in\mathcal{V}, t'\in \{0,1,2,\cdots\}\}$. If the walker is at the vertex $i$, it steps into a neighbor via the activated edge randomly. Consider the probability that the walker steps from $i$ to $j$ with the condition that the edge $(i, j)$ is activated (with probability $q_0$ as mentioned previously). At time $t$, the probability of finding $d$ activated edges around the vertex $i$ ($1\leq d\leq k_i$) is 
\begin{equation}
P[\sum_{h\in\mathcal{G}}x_{ih}(t)=d\vert x_{ij}=1]=q_0\frac{(k_i-1)!}{(d-1)!(k_i-d)}q_0^{d-1}(1-q_0)^{k_i-d}.
\end{equation}
The probability that the random walker steps into $j$ is $1/d$. If the edge $(i, j)$ is not activated or does not exist, the probability of a random walk from $i$ to $j$ is $0$. Therefore, the probability for a random walk from $i$ to $j$ is 
\begin{equation}
\begin{aligned}
l_{ij}=&\sum_{d=1}^{k_i}\frac{1}{d}P[\sum_{h\in\mathcal{G}}x_{ih}(t)=d\vert x_{ij}=1]\\
&=\sum_{d=1}^{k_i}\frac{1}{k_i}\frac{k_i!}{d!(k_i-d)!}q_0^d(1-q_0)^{k_i-d}\\
&=\frac{1-(1-q_0)^{k_i}}{k_i}. 
\end{aligned}
\end{equation}
We have mentioned that a vertex can be isolated in activated subgraph $\mathcal{G}_A(t)$. Since the nature of probability, we define that the walker steps into itself when a Poisson event for the random walk occurs if the focal vertex has no activated edge out, with the probability $(1-q_0)^{k_i}$. This directly leads to the conclusion in Eq.~\ref{eq: one step random walk}. 

Evidently, the process $R'(t')$ has a unique stationary distribution $P[R'(t')=i]$ because this Markov chain is irreducible and periodic, which is equivalent to the stationary distribution of $R(t)$. Since the random walk process is time reversible, we have
\begin{equation}
P[R'(t')=i]\frac{1-(1-q_0)^{k_i}}{k_i}=P[R'(t')=j]\frac{1-(1-q_0)^{k_j}}{k_j}=c, 
\end{equation}
where $c$ is an auxiliary constant. Using $\sum_{j\in\mathcal{G}}P(R'(t')=j)=1$, we have $c=1/(\sum_{j\in\mathcal{G}}\frac{k_j}{1-(1-q_0)^{k_j}})$. Therefore, the stationary distribution of $R'(t')$ is
\begin{equation}
P[R'(t')=i]=\frac{k_i}{[1-(1-q_0)^{k_i}] \sum_{j\in\mathcal{G}}\frac{k_j}{1-(1-q_0)^{k_j}}}, 
\end{equation}
which directly results in Eq.~\ref{eq: random walk stationary} for $R(t)$. For isothermal graphs where each vertex has the same degree, since there are $N$ vertices, this distribution degenerates into $P[R(t)=1/N]$ as a uniform distribution. 
\subsection{Game, Payoff, and Replacement}\label{Methods: 3}
\hspace{1.5em}We employ the donation game with $b>c>0$, i.e., the prisoner's dilemma (PD), to study the evolution of cooperation in networks with the stochastic switching topology. The payoff matrix of the PD is
\begin{equation}
\begin{array} {cc}
& \begin{array}{cc}  C&D\end{array}\\
\begin{array}{c} C\\D\end{array}&
\left[\begin{array}{cc}
		b-c & -c \\
		b & 0 \\\end{array}\right]
\end{array}
\end{equation}
Consider a local game with two players, a cooperator receives $b-c$ or $-c$ when interacting with a cooperator or defector respectively. A defector obtains the temptation $b$ against a cooperator. Interaction between two defectors yields no payoff to each other. We denote the strategy of the population as a vector $\mathbf{S}(t)=(s_0(t), s_1(t), \cdots, s_N(t)), s_i(t)\in\{0,1\}$, where $s_i(t)=0$ (\textit{resp.} $s_i(t)=1$) indicates the individual $i$'s strategy is defection (\textit{resp.} cooperation). Therefore, $i$'s payoff at time $t$ is the sum of the game payoff from all activated neighbors
\begin{equation}
\begin{aligned}
&\pi_i(t)=\sum_{j\in\mathcal{G}}x_{ij}(t)[(b-c)s_i(t)s_j(t)+(-c)s_i(t)(1-s_j(t))\\
&+b(1-s_i(t))s_j(t)]=\sum_{j\in\mathcal{G}}x_{ij}(t)[-cs_i(t)+bs_j(t)]. 
\end{aligned}
\end{equation}
This payoff is transformed into fecundity by the exponential function $F_i(t)=\exp(w\pi_i(t))$. Under weak selection, using Taylor's formula, we obtain
\begin{equation}
F_i(t)=1+w\pi_i(t)+O(w^2). 
\end{equation}
The replacement occurs by the Poisson process with the rate of $1$ following the death-birth updating. If there exist activated edges around $i$, $i$'s neighbors spread their strategy through the activated links to $i$ with the probability proportional to fecundity. If all edges are dormant around $i$, it keeps its original strategy and no replacement from neighbors occurs. Therefore, a neighbor $j$ spreads its strategy to $i$ with the probability
\begin{equation}
P[s_i(t)\leftarrow s_j(t)]=[1-(1-q_0)^{k_i}]\frac{x_{ij}(t)F_j(t)}{\sum_{h\in\mathcal{G}}x_{ih}(t)F_h(t)}. 
\end{equation}
Individual $i$ keeps its original strategy with the trapped probability
\begin{equation}
P[s_i(t)\leftarrow s_i(t)]=(1-q_0)^{k_i}. 
\end{equation}
\enlargethispage{20pt}
\section*{Acknowledgement}
Z.Z. and M.F. are supported by the National Natural Science Foundation of China (NSFC) (Grant No. 62206230) and the Natural Science Foundation of Chongqing (Grant No. CSTB2023NSCQ-MSX0064). M.P. was supported by the Slovenian Research and Innovation Agency (Javna agencija za znanstvenoraziskovalno in inovacijsko dejavnost Republike Slovenije) (Grant Nos. P1-0403 and N1-0232).
\section*{Data Access}
The authors made use of the published data sets in \cite{jeong2001lethality,cho2014wormnet, nr-aaai15, bigbrain,ahmed2010time}. Our source files can be found in \url{https://github.com/TheUsernameNULL/Bursty-Switching-in-Complex-Networks-Code}.
\vskip2pc
\bibliography{sn-bibliography}
\end{document}